\documentclass[smallextended]{svjour3}       % onecolumn (second format)
\smartqed  % flush right qed marks, e.g. at end of proof
\usepackage{graphicx}
\usepackage{amssymb}
\usepackage{amsmath}
\usepackage[colorlinks = true,
linkcolor = blue,
urlcolor  = blue,
citecolor = blue,
anchorcolor = blue]{hyperref}% add hypertext capabilities
\usepackage{xcolor}
\usepackage{subfig}
\renewcommand{\figurename}{\textbf{Fig.}}
\renewcommand{\tablename}{\textbf{Table}}
\usepackage{graphicx}% Include figure files
\usepackage{dcolumn}% Align table columns on decimal point
\usepackage{bm}% bold math
\usepackage{epstopdf}
\usepackage{epsfig}
\DeclareMathOperator{\sech}{sech}

\newtheorem{thm}{Theorem}
\begin{document}
	
	\title{Nonlinear states of the conservative complex Swift-Hohenberg equation}
	
	
	\author{R. Kusdiantara \and H. Susanto}
	
	
	\institute{
		R. Kusdiantara \at
		Industrial and Financial Mathematics Research Group, Faculty of Mathematics and Natural Sciences, Institut Teknologi Bandung, Jl. Ganesha 10, Bandung, 40132, Indonesia\\
		\email{rudy\_kusdiantara@itb.ac.id}           %  \\
		\and
		R. Kusdiantara \at
		Centre of Mathematical Modelling and Simulation, Institut Teknologi Bandung, Jl. Ganesha 10, Bandung, 40132, Indonesia\\
		%		\email{rudy\_kusdiantara@itb.ac.id}\\
		\and
		H. Susanto \at
		Department of Mathematics, Khalifa University of Science and Technology, PO Box 127788, Abu Dhabi, United Arab Emirates\\
		\email{hadi.susanto@ku.ac.ae}
		\and          
		H. Susanto \at 
		Department of Mathematics, %Faculty of Mathematics and Natural Sciences, 
		Universitas Indonesia, Gedung D Lt.\ 2 FMIPA Kampus UI, Depok, 16424, Indonesia
	}
	
	\date{Received: date / Accepted: date}

	\maketitle
	
	\begin{abstract}
		We consider the conservative complex Swift-Hohenberg equation, which belongs to the family of nonlinear fourth-order dispersive Schr\"odinger equations. In contrast to the well-studied one-dimensional dissipative Swift-Hohenberg equation, the complex variant introduces a wide array of largely unexplored solutions. Our study provides a fundamental step in understanding the complex characteristics of this equation, particularly for typical classes of solutions-uniform, periodic, and localized states-and their relationship with the original dissipative model. Our findings reveal significant differences between the two models. For instance, uniform solutions in the conservative model are inherently unstable, and periodic solutions are generally unstable except within a narrow parameter interval that supports multiple localized states. Furthermore, we establish a generalized Vakhitov-Kolokolov criterion to determine the stability of localized states in the conservative equation and relate it to the stability properties of the dissipative counterpart.
	\end{abstract}
	
	\section{Introduction}
	The Swift-Hohenberg equation (SHE) \cite{swift1977hydrodynamic}-or more precisely, the Turing-Swift-Hohenberg equation, as a similar equation with slightly different nonlinearity was derived by Alan Turing over 50 years before the influential work by Swift and Hohenberg \cite{dawes2016after}-is one of the simplest and most canonical models for pattern formation. It has been widely applied across various physical contexts, including ecological systems \cite{tlidi2008vegetation}, neuronal populations \cite{hutt2007generalization,laing2002multiple}, and cellular buckling \cite{hunt1989structural,thompson1988spatial}. Originally derived to study the nonequilibrium transition from a uniform to a nonuniform convecting state, numerical studies have revealed its rich spatiotemporal dynamics, establishing it as a foundational model for pattern formation \cite{cross1993pattern}. Localized structures separating domains in higher-dimensional SHEs have been extensively analyzed, both theoretically and numerically \cite{staliunas1998dynamics,taranenko1998pattern}, highlighting the equation's versatility in capturing diverse pattern-forming behaviors.
	
	Extending the model to include complex-valued parameters yields the complex SHE (CSHE), enabling the study of oscillatory instabilities and richer spatiotemporal dynamics not captured by the real-valued SHE \cite{khairudin2016stability}. First introduced in \cite{malomed1984nonlinear}, the CSHE resembles the Ginzburg-Landau equation \cite{staliunas1993laser,aranson2002world,garcia2012complex} but includes a fourth-order spatial derivative term. It has been used to model physical systems such as optical parametric oscillators \cite{sanchez1997generalized,longhi1996swift,staliunas1997nonlinear}, photorefractive oscillators \cite{staliunas1995analogy}, and class A and C lasers \cite{lega1994swift,lega1995universal,mercier2002derivation}. Pedrosa et al.\ \cite{pedrosa2008numerical} numerically validated the CSHE as an asymptotic reduction of the Maxwell-Bloch equations for class C lasers with small detuning. The fourth-order derivative introduces spectral filtering, making the model more realistic for systems with complex spectral responses \cite{soto2002composite}.
	
	The CSHE exhibits a vast array of nontrivial solutions and dynamics \cite{staliunas1997nonlinear,soto2002composite,sakaguchi1996stable,aranson1995domain,gelens2010coarsening}. To systematically analyze these, it is often necessary to relate them to well-known characteristics in limiting scenarios. Soto-Crespo and Akhmediev \cite{soto2002composite} explored a subset of CSHE solutions by connecting them to the complex Ginzburg-Landau equation, highlighting the challenges posed by the fourth-derivative term as a singular perturbation. This work provides a foundational step in solution classification by examining the stability differences between the dissipative SHE and the conservative CSHE, particularly focusing on how the stability of solutions changes between the two.
	
	The conservative CSHE, previously studied in \cite{knobloch2014stability}, belongs to the class of nonlinear Schr\"odinger equations with a fourth-order derivative term \cite{blow1989theoretical,karlsson1994soliton,karpman1991influence,karpman2000stability,ilan2002self}. While its physical relevance has been discussed in \cite{ilan2002self}, recent interest has been spurred by the discovery of pure-quartic solitons in experimental media \cite{blanco2016pure,runge2020pure}. These solutions have been numerically analyzed \cite{tam2019stationary}, with studies on their formation from Gaussian initial conditions \cite{tam2020generalized} and the existence and stability of multi-pulse configurations \cite{parker2021multi}. Despite the long history of fourth-order dispersive nonlinear Schr\"odinger equations, our work presents new results by focusing on standing waves analogous to those in the dissipative SHE studied by \cite{burke2007snakes}, which exhibit phenomena such as homoclinic snaking \cite{woods1999heteroclinic}. These solutions have not been previously studied in the context of Schr\"odinger equations. Davydova and Zaliznyak \cite{davydova2001schrodinger} examined a similar equation with quadratic and quartic dispersions and cubic-quintic nonlinearity but focused solely on single-hump solutions. Additionally, we demonstrate that even uniform and spatially periodic states exhibit distinct stability characteristics. {We also want to refer the reader to the work of Ponedel and Knobloch \cite{ponedel2018gap} on the cubic-quintic Gross-Pitaevskii equation with a quadratic dispersion and an external potential $V(x)$ that yields forced snaking and gap solitons. While our study explores states in a system with a quartic dispersion, our results can be extended to the quadratic dispersion case.}
	
	Our paper is structured as follows. Section \ref{sec2} introduces the model and numerical methods. Section \ref{sec3} examines uniform and periodic solutions, including their instability dynamics. Localized states are analyzed in Section \ref{sec4}, and the relationship between CSHE states and their dispersive counterparts is established through a Vakhitov-Kolokolov equation in Section \ref{sec5}. Finally, we conclude in Section \ref{conc}.
	
	\section{Main model}
	\label{sec2}
	
	The dissipative SHE is given by \cite{burke2007snakes}
	\begin{equation}
		\psi_t = -(1+\partial_{xx})^2\psi - \omega\psi + b_3\psi^3 - \psi^5,
		\label{dshe}
	\end{equation}
	where $\psi(x,t)$ represents the real-valued field, and $x$ and $t$ are the spatial and temporal variables, respectively. The parameters $r$ and $b_3$ are also real. However, {in this report, we consider the conservative CSHE given by:
		\begin{equation}
			i\psi_t = -(1+\partial_{xx})^2\psi + b_3|\psi|^2\psi - |\psi|^4\psi,
			\label{gov}
		\end{equation}
		where $\psi(x,t)$ now becomes complex-valued. The parameter $b_3$ is still real; unless otherwise mentioned, we set $b_3 = 2$. The equation can be derived from the action:
		\begin{equation}
			S = \int \mathcal{L}\, dx\, dt,
		\end{equation}
		with the Lagrangian density:	
		\[
		\mathcal{L} = \frac{i}{2} \left( \psi \psi^*_t - \psi^* \psi_t \right) - |\psi|^2 + 2|\psi_x|^2 - |\psi_{xx}|^2 + \frac{b_3}{2}|\psi|^4 - \frac{1}{2}|\psi|^6.
		\]
		The power
		\begin{equation}
			P = \int_{-\infty}^\infty |\psi|^2 \, dx
			\label{P}
		\end{equation}
		is conserved in this framework.}
	
	Although systems \eqref{dshe} and \eqref{gov} appear similar, their characteristics are fundamentally different, as one is dissipative and the other is Hamiltonian. The conservative system described by \eqref{gov} supports standing wave solutions characterized by phase invariance and energy conservation, whereas the dissipative counterpart features steady states where energy dissipation plays a crucial role. This distinction is significant when analyzing stability properties, as the same solution may exhibit different stability behaviors in the two systems, as we will discuss below. Our study aims to bridge this gap by analyzing and contrasting the stability characteristics of steady states in the two systems.
	
	To study the standing waves of Eq.\ \eqref{gov}, we transform to a rotating reference frame by considering solutions of the form \(\psi(x,t) = e^{-i\omega t}u(x)\), where \(u(x)\) is a real-valued function. Substituting this ansatz into Eq.\ \eqref{gov} yields the stationary equation:
	\begin{equation}
		-\omega u - (1 + \partial_{xx})^2u + b_3u^3 - u^5 = 0.
		\label{gov2}
	\end{equation}
	This equation describes standing wave solutions and is identical to the steady-state equation of \eqref{dshe}, whose bifurcations and localized solutions were analyzed in detail in \cite{burke2007snakes}. Here, we extend these studies by focusing on the stability properties of such solutions within the framework of the conservative CSHE. To analyze the stability of these states, we introduce a perturbed solution of the form:
	\[
	\psi(x,t) = e^{-i\omega t} \left[ u(x) + \left( r(x) + is(x) \right)e^{\lambda t} + \left( r(x)^* + is(x)^* \right)e^{\lambda^* t} \right],
	\]
	where \(r(x)\) and \(s(x)\) represent small perturbations around the stationary state \(u(x)\), and \(\lambda\) denotes the eigenvalues associated with these perturbations. Substituting this ansatz into Eq.\ \eqref{gov} and linearizing about \(r(x)\) and \(s(x)\) leads to the eigenvalue problem:
	\begin{equation}
		\lambda
		\begin{pmatrix}
			r \\ s
		\end{pmatrix}
		= \mathcal{J}\mathcal{L}
		\begin{pmatrix}
			r \\ s
		\end{pmatrix},
		\label{stab}
	\end{equation}
	where the operators \(\mathcal{J}\) and \(\mathcal{L}\) are defined as:
	\[
	\mathcal{J} = \begin{pmatrix}
		0 & 1 \\
		-1 & 0
	\end{pmatrix}, \quad
	\mathcal{L} = \begin{pmatrix}
		L_+ & 0 \\
		0 & L_-
	\end{pmatrix}.
	\]
	The operators \(L_+\) and \(L_-\) are given by:
	\begin{equation}
		\begin{split}
			L_- &= -\omega - (1 + \partial_{xx})^2 + b_3u^2 - u^4, \\
			L_+ &= -\omega - (1 + \partial_{xx})^2 + 3b_3u^2 - 5u^4.
		\end{split}
	\end{equation}
	Here, \(L_-\) and \(L_+\) represent the linear operators governing perturbations in the imaginary and real parts of the solution, respectively.
	
	The stability of the state \(u(x)\) in the context of the CSHE \eqref{gov} is determined by the spectrum of the eigenvalues \(\lambda\) obtained from Eq.\ \eqref{stab}. Specifically, a solution is linearly unstable if there exists at least one eigenvalue with \(\text{Re}(\lambda) > 0\). However, the symplectic structure introduced by the operator \(\mathcal{J}\) imposes symmetry properties on the eigenvalue spectrum: if \(\lambda\) is an eigenvalue, then \(-\lambda\) and \(\pm\lambda^*\) are also eigenvalues. Consequently, a solution is linearly stable only when all eigenvalues are purely imaginary. In contrast, for the dissipative SHE \eqref{dshe}, the stability of \(u(x)\) is determined solely by the spectrum of the operator \(L_+\), i.e., it is linearly stable only when all eigenvalues have non-positive real parts. Thus, the stability of the same solution can differ between the conservative and dissipative contexts. Identifying and understanding these differences is a central focus of this paper.
	
	We study the CSHE numerically using the following procedure. First, we solve the stationary equation \eqref{gov2} using Newton's method, approximating spatial derivatives via a spectral method to achieve high accuracy. Once a solution is obtained, its stability is evaluated by solving the corresponding discretized eigenvalue problem \eqref{stab}. For solutions identified as unstable, we investigate their time dynamics by directly solving the governing equation \eqref{gov} using a fourth-order Runge-Kutta method for numerical integration. This approach enables us to observe how instabilities manifest and evolve over time.
	
	\begin{figure*}[htbp!]
		\centering
		\includegraphics[scale=0.43]{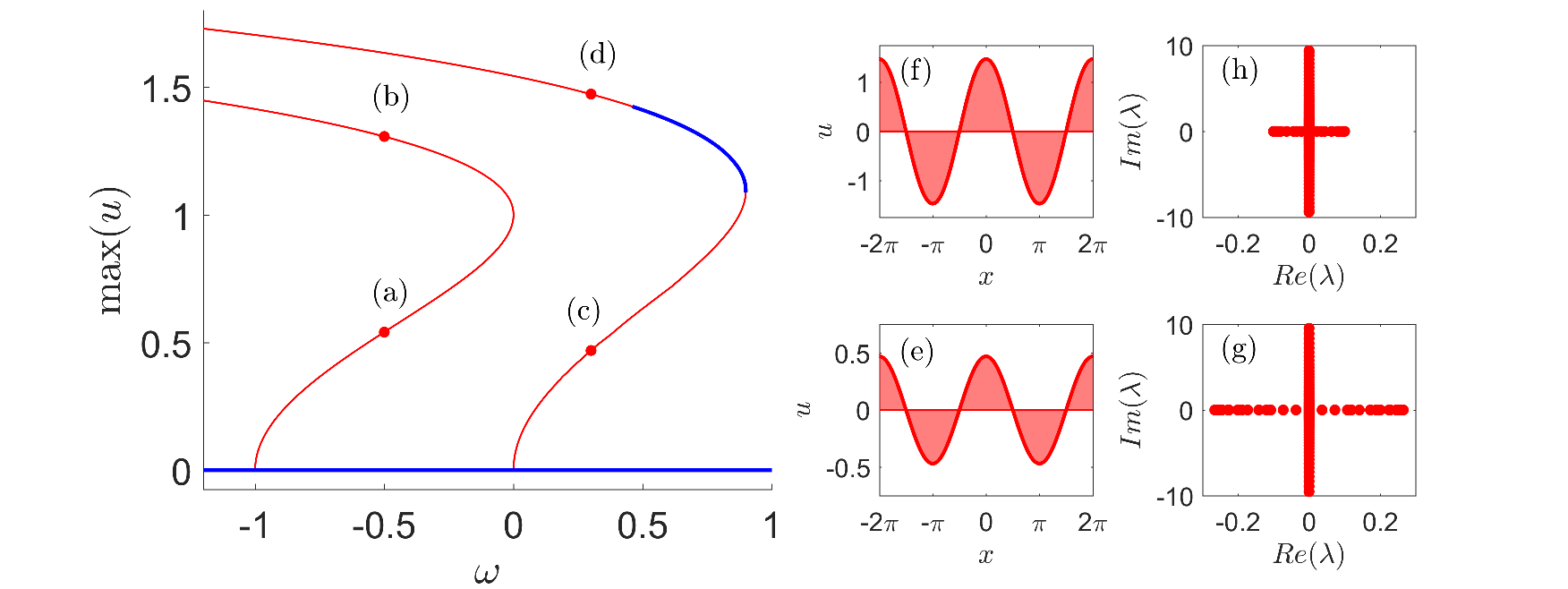}
		\caption{{The left panel presents the bifurcation diagram of the uniform and periodic states of the governing equation \eqref{gov}, with uniform states bifurcating from $\omega = -1$ and periodic states from $\omega = 0$. Stable and unstable branches are denoted by solid blue and dashed red lines. Key points marked with letters (a)--(d) correspond to solution profiles and stability analyses discussed below. Panels (e) and (f) depict the spatial profiles of periodic solutions at points (c) and (d) on the bifurcation diagram, both for $\omega = 0.3$. Panels (g) and (h) illustrate the corresponding spectrum of the periodic solutions shown in panels (e) and (f), respectively, in the complex plane. The presence of eigenvalues with positive real parts determines instability. 
			}
		}		
		\label{fig:unisol_persol}
	\end{figure*}
	
	\begin{figure*}[htbp!]
		\centering
		\subfloat[]{\includegraphics[scale=0.39]{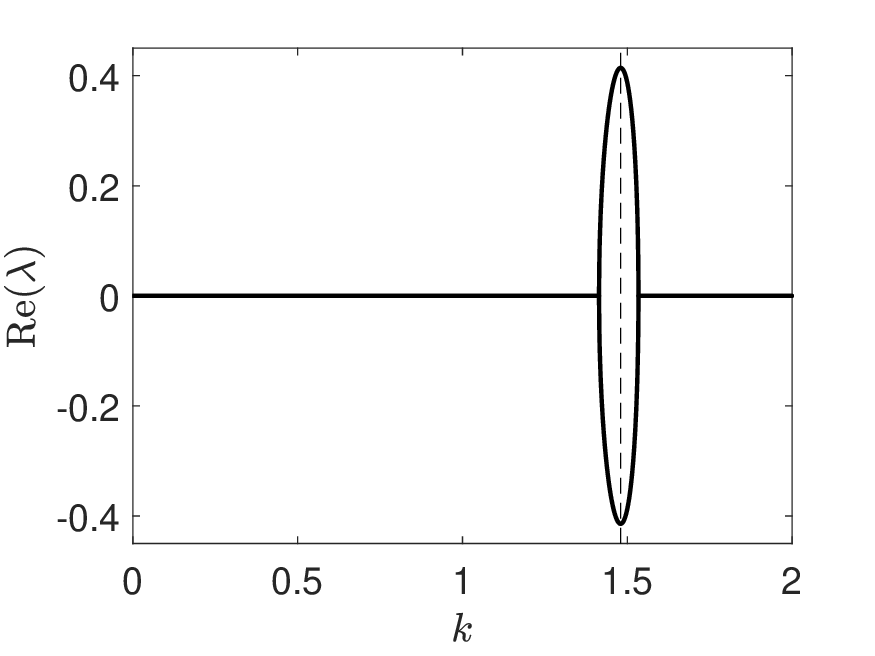}\label{subfig:disper_real}}\,\,
		\subfloat[]{\includegraphics[scale=0.39]{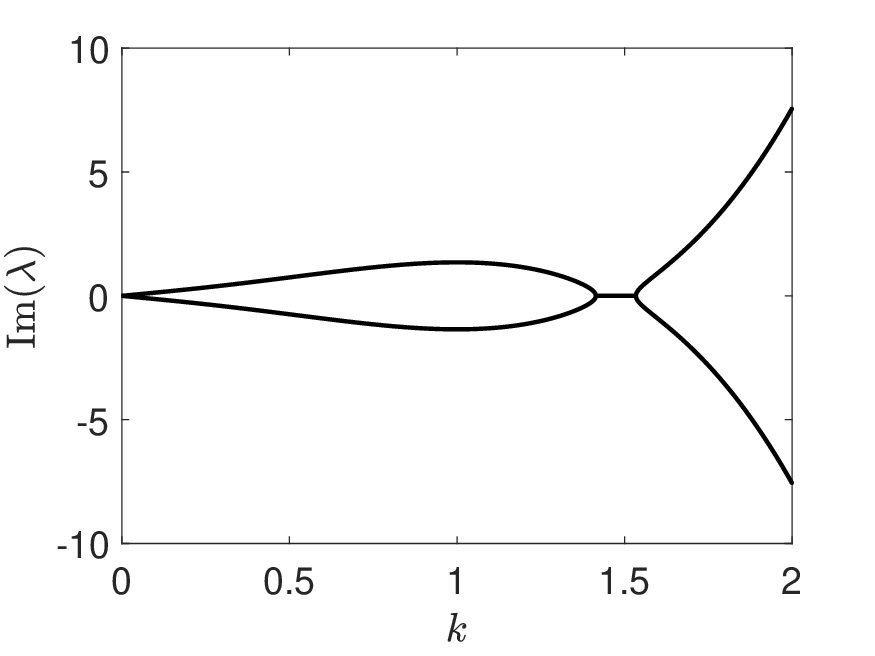}\label{subfig:disper_imag}}\,\,
		\caption{The dispersion relation of the uniform solution $u_-$ at point (a) in Fig.\ \ref{fig:unisol_persol}. Panel (a) shows the spectrum real part, $\text{Re}(\lambda)$, as a function of the wavenumber $k$, which determines the growth or decay rate of perturbations. {The vertical dashed line shows the critical wavenumber \( k_c \), where the growth rate is maximum.} Panel (b) displays the imaginary part, $\text{Im}(\lambda)$, representing the oscillatory behavior of perturbations. $u_-$ is modulationally unstable because $\text{Re}(\lambda)>0$ within a certain interval of $k$.
		}	
		\label{fig:disper}
	\end{figure*}
	
	\section{Uniform and periodic solutions}
	\label{sec3}

	\begin{figure}[tbhp!]
		\centering
		\subfloat[]{\includegraphics[scale=0.39]{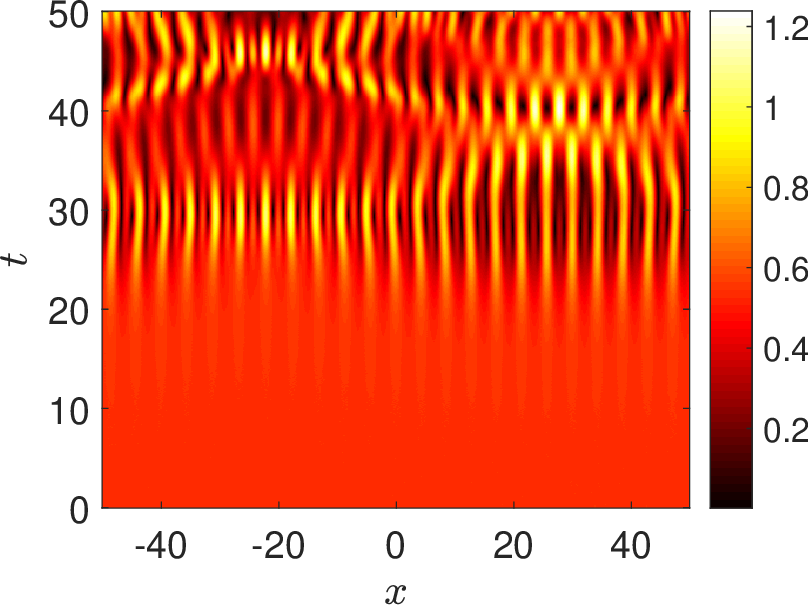}\label{subfig:unisol_lower_c}}\,\,
		\subfloat[]{\includegraphics[scale=0.39]{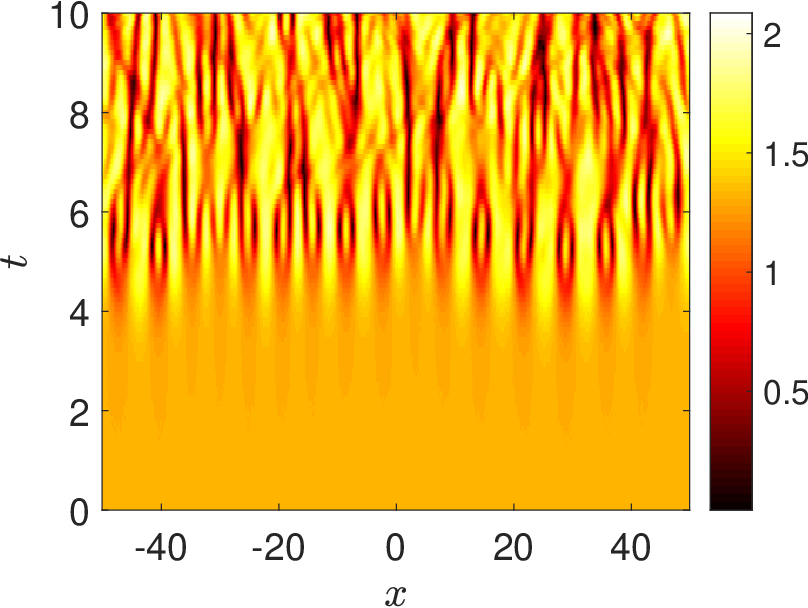}\label{subfig:unisol_upper_d}}\\
		\subfloat[]{\includegraphics[scale=0.39]{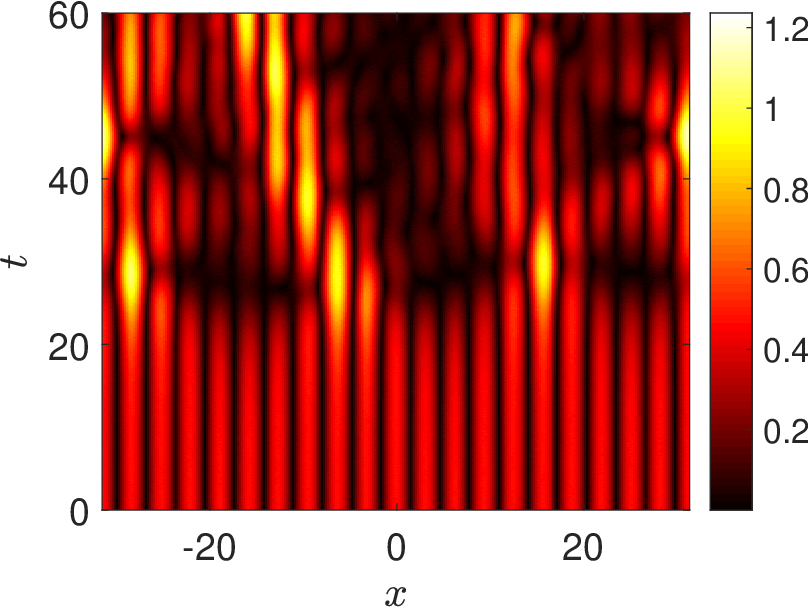}\label{subfig:persol_lower_a}}\,\,
		\subfloat[]{\includegraphics[scale=0.39]{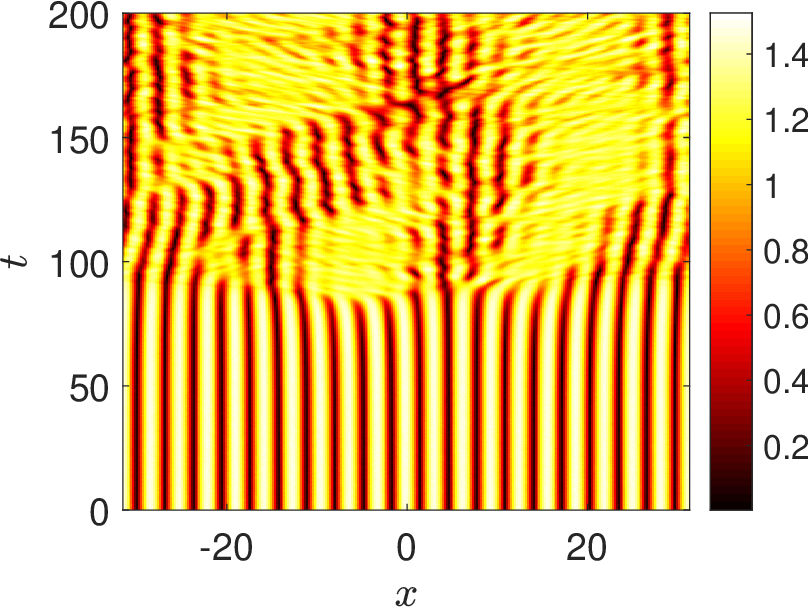}\label{subfig:persol_upper_b}}
		%		\caption{Time-dynamics of the unstable uniform and periodic  solutions that are indicated as points (a)-(d) in Fig.\ \ref{fig:unisol_persol}.}
		\caption{{The time evolution of unstable uniform and periodic solutions, corresponding to points (a)--(d) in Fig.\ \ref{fig:unisol_persol}. Panels (a) and (b) illustrate the evolution of the lower and upper uniform states, respectively, developing spatially and temporally varying patterns over time. Similarly, panels (c) and (d) capture the dynamics of the lower and upper periodic solutions, where instabilities result in either localized periodic patches or plateau-like structures.}}
		\label{fig:time_dynamics_unisol_persol}
	\end{figure}
	
	The conservative CSHE \eqref{gov2} admits the same uniform and periodic solutions as those of the SHE \eqref{dshe}, which were extensively studied in \cite{burke2007snakes}. The uniform solutions are given by:
	\begin{equation}
		u_0 = 0 \quad \text{and} \quad u_{\pm}^2 = \frac{1}{2} \left( b_3 \pm \sqrt{b_3^2 - 4 \left( \omega + 1 \right)} \right).
		\label{eq:unisol_sol}
	\end{equation}
	Here, \(u_0 = 0\) represents the trivial solution, while \(u_+\) and \(u_-\) correspond to the non-trivial upper and lower uniform states, respectively. Fig.\ \ref{fig:unisol_persol} shows the solution bifurcation diagram. The stability of these solutions is determined by analyzing the eigenvalues \(\lambda\) obtained from the stability equations \eqref{stab}. For the uniform states $u_j$, $j=0,\pm$, this analysis involves substituting perturbations of the form \(r(x), s(x) \sim e^{ikx}\) into the linearized equations, leading to the following dispersion relation:
	\begin{equation}
		\lambda(k,u_j) = \pm \sqrt{-L_1 L_2},\label{add2}
	\end{equation}
	where the operators \(L_1\) and \(L_2\) are given by:
	\begin{equation}
		\begin{aligned}
			L_1 &= -\omega - (1 - k^2)^2 + b_3u_j^2 - u_j^4, \\
			L_2 &= -\omega - (1 - k^2)^2 + 3b_3u_j^2 - 5u_j^4.
		\end{aligned}
	\end{equation}
	%	Analyzing Eq.\ \eqref{add2}, we obtain that the trivial uniform state \( u_0 = 0 \) is always stable because \( -L_1L_2 \leq 0 \) for all wavenumbers \( k \). The non-trivial uniform states \( u_{\pm} \) are modulationally unstable because \( -L_1L_2 > 0 \) within a certain interval of \( k \) that depends on the parameter values \( b_3 \) and \( \omega \). In Fig.\ \ref{fig:disper}, we plot the dispersion relation of the lower uniform solution indicated as point (a) in Fig.\ \ref{fig:unisol_persol}. The figure shows that $u_-$ is unstable to periodic perturbations with certain wavenumbers $k$. 
	%	\textcolor{red}{The critical wave number $k_c$, attains at 
		%	$$
		%	k_c=\left\{\begin{array}{ccc}
			%		1&,& \text{for } u_+\\
			%		 \left(1+\left(-\omega+2b_3u_-^2-3u_-^4\right)^{\frac{1}{2}}\right)^{\frac{1}{2}}&,& \text{for } u_-
			%	\end{array}
		%	\right.
		%	$$}
	%	The stability results are summarized and depicted in Fig.\ \ref{fig:unisol_persol} coded in colors. These characteristics in the conservative CSHE \eqref{gov} differ from those in the SHE \eqref{dshe}, where the stability can change: the zero solution may lose stability, the upper non-zero solution $u_+$ is typically stable, and the lower one $u_-$ remains unstable, as shown in \cite[Fig.\ 1]{burke2007snakes}. 
	{Analyzing Eq.\ \eqref{add2}, we find that the trivial uniform state \( u_0 = 0 \) is always stable, as \( -L_1L_2 \leq 0 \) for all wavenumbers \( k \). In contrast, the non-trivial uniform states \( u_{\pm} \) exhibit modulational instability because \( -L_1L_2 > 0 \) within a certain range of \( k \). The specific interval of instability depends on the parameter values \( b_3 \) and \( \omega \).
		To illustrate this, Fig.\ \ref{fig:disper} presents the dispersion relation for the lower uniform solution \( u_- \), as indicated by point (a) in Fig.\ \ref{fig:unisol_persol}. The real part of the eigenvalues \( \text{Re}(\lambda) \), shown in panel (a), reveals that \( u_- \) is unstable to periodic perturbations with certain wavenumbers \( k \). This instability is further characterized by the imaginary part \( \text{Im}(\lambda) \), displayed in panel (b), which highlights the oscillatory dynamics of these perturbations.
		The critical wavenumber \( k_c \), at which the instability growth rate is maximal, is given by:
		\[
		k_c =
		\begin{cases} 
			1, & \text{for } u_+ \\
			\left(1 + \sqrt{-\omega + 2b_3u_-^2 - 3u_-^4}\right)^{1/2}, & \text{for } u_-
		\end{cases}
		\]
		This expression shows that \( k_c \) depends on the system parameters and the considered solution branch.
		The stability results for the uniform states are summarized in Fig.\ \ref{fig:unisol_persol}, with stability and instability coded by colors. These results demonstrate notable differences between the conservative complex Swift-Hohenberg equation (CSHE) \eqref{gov} and the dissipative Swift-Hohenberg equation (SHE) \eqref{dshe}. In the SHE, stability can vary: the zero solution may lose stability, the upper non-trivial solution \( u_+ \) is generally stable, and the lower one \( u_- \) remains unstable, as observed in \cite[Fig.\ 1]{burke2007snakes}. The structure of the corresponding eigenvalue problem causes the difference. %However, in the conservative CSHE, the stability characteristics are fundamentally altered due to the symplectic structure of the system.
	}
	
	Following the analysis of \cite{burke2007snakes}, Eq.\ \eqref{gov2} also admits spatially periodic solutions that bifurcate from the uniform solution at $\omega = 0$. The bifurcation diagram for these periodic solutions is illustrated in Fig.\ \ref{fig:unisol_persol}. Stability analysis reveals that periodic states are generally unstable, except within a narrow interval of the bifurcation parameter $\omega \in \left[0.4604, 0.8991\right]$. Specifically, as shown in Fig.\ \ref{fig:unisol_persol}(e)-(h), the solutions labeled (c) and (d) in the bifurcation diagram (left panel) exhibit instability due to bands of eigenvalues on the real axis in the complex spectrum. This behavior contrasts sharply with the SHE \eqref{dshe}, where the upper periodic solution remains stable across all bifurcation parameter values, as demonstrated in \cite[Fig.\ 2]{burke2007snakes}.
	
	The time dynamics of unstable solutions offer further insight into their behavior. For the spatially homogeneous and periodic solutions labeled (a)--(d) in Fig.\ \ref{fig:unisol_persol}, Fig.\ \ref{fig:time_dynamics_unisol_persol} illustrates their characteristic instability dynamics. Specifically, Figs.\ \ref{subfig:unisol_lower_c} and \ref{subfig:unisol_upper_d} show the time evolution of the lower and upper uniform states, respectively, at $\omega = -0.5$. Over time, the homogeneous solution transitions into a state exhibiting both spatial and temporal periodicity. Similarly, for spatially periodic states at $\omega = 0.3$, Figs.\ \ref{subfig:persol_lower_a} and \ref{subfig:persol_upper_b} demonstrate how the instability manifests, leading to either localized patches of periodic states or plateau-like structures. These simulations elucidate the distinct instability mechanisms governing the uniform and periodic solutions along the unstable branches of the bifurcation diagram.
	
	\begin{figure}[t!]
		\centering
		\subfloat[]{\includegraphics[scale=0.39]{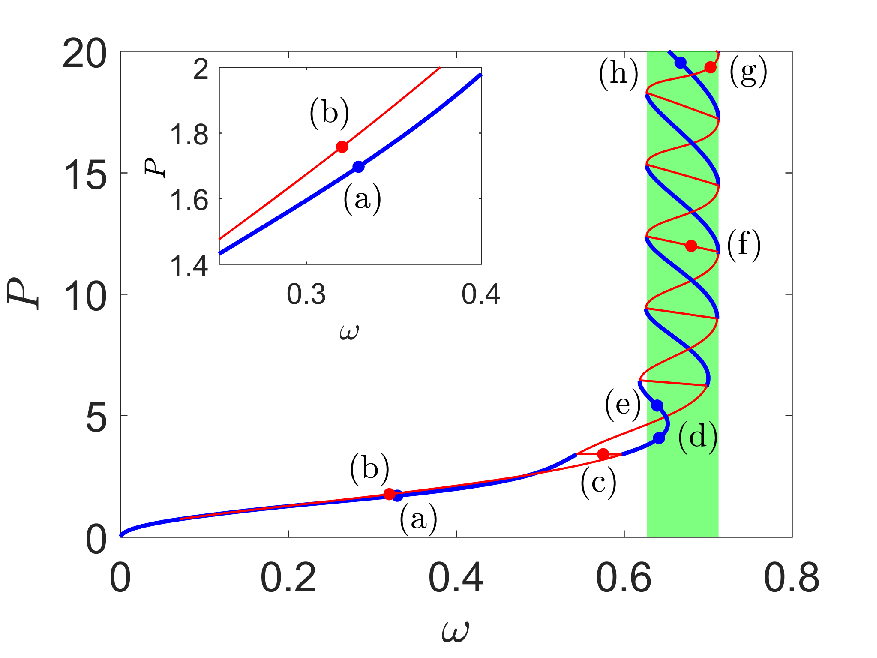}	\label{fig:snakes}}
		\subfloat[]{\includegraphics[scale=0.39]{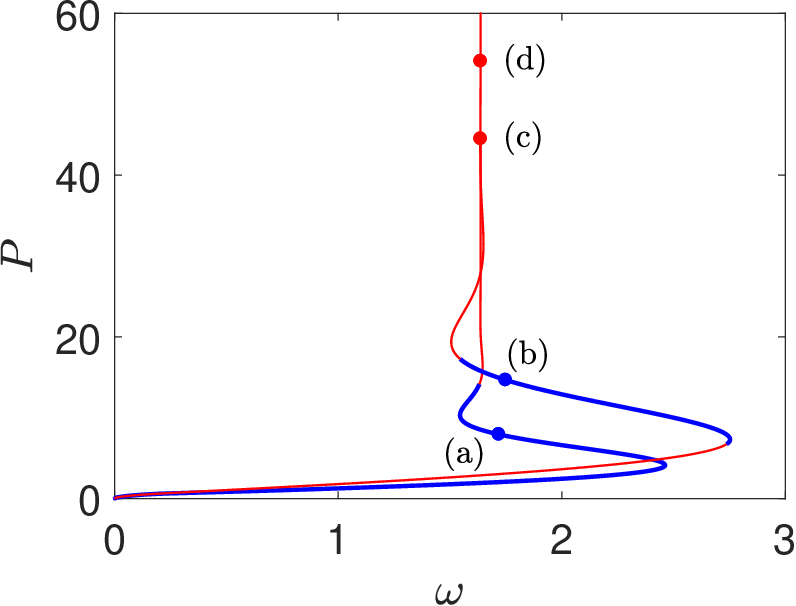}\label{fig:snakes_b3}}
		\caption{Bifurcation diagrams of localized solutions with $\phi=0,\,\pi$ and $\phi=\pi/2,\,3\pi/2$ and their stability for the conservative CSHE. The inset zooms in on the neighbourhood of point (a) and (b). %In this case, the localized states are formed by a front and a back that correspond to a heteroclinic connection between the zero state $u_0$ and a periodic state. 
			{The corresponding solutions of points (a)--(h) with their spectrum depicted are presented  in Fig.\ \ref{fig:profiles_eigenvalues}.} In panel (b), $b_3 = 3.75$, where there is no snaking characteristics in the bifurcation diagram of the localized solutions. %In this case, the front and back correspond to a heteroclinic connection between two uniform states, i.e., the zero state $u_0$ and the upper non-zero state $u_+$. 
			The corresponding solutions of points (a)--(d) and their spectrum are presented in Fig.\ \ref{fig:prof_eigen_b3}. Blue thick and red thin lines indicate stable and unstable solutions, respectively. 
		}
		\label{fig:add1}	
	\end{figure}
	
	\begin{figure}[tbhp!]
		\centering
		\subfloat[]{\includegraphics[scale=0.19]{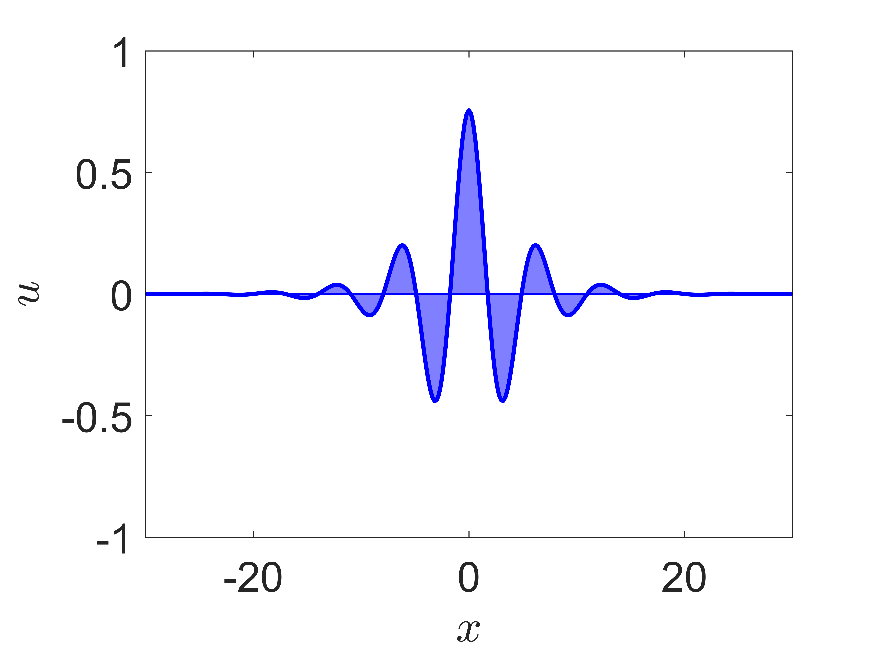}\label{subfig:prof_a}}\,\,
		\subfloat[]{\includegraphics[scale=0.19]{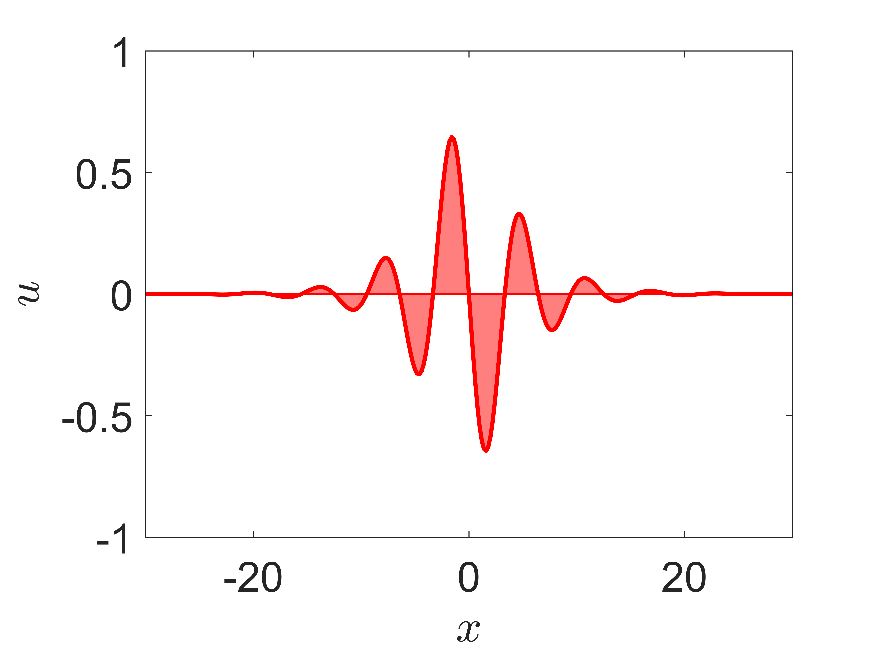}\label{subfig:prof_b}}
		\subfloat[]{\includegraphics[scale=0.19]{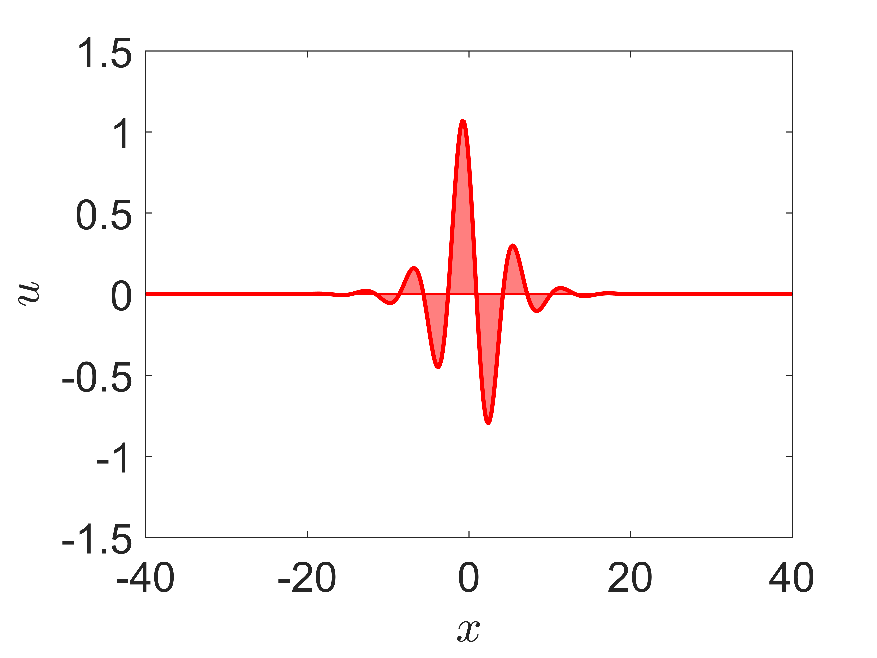}\label{subfig:prof_c}}\,\,
		\subfloat[]{\includegraphics[scale=0.19]{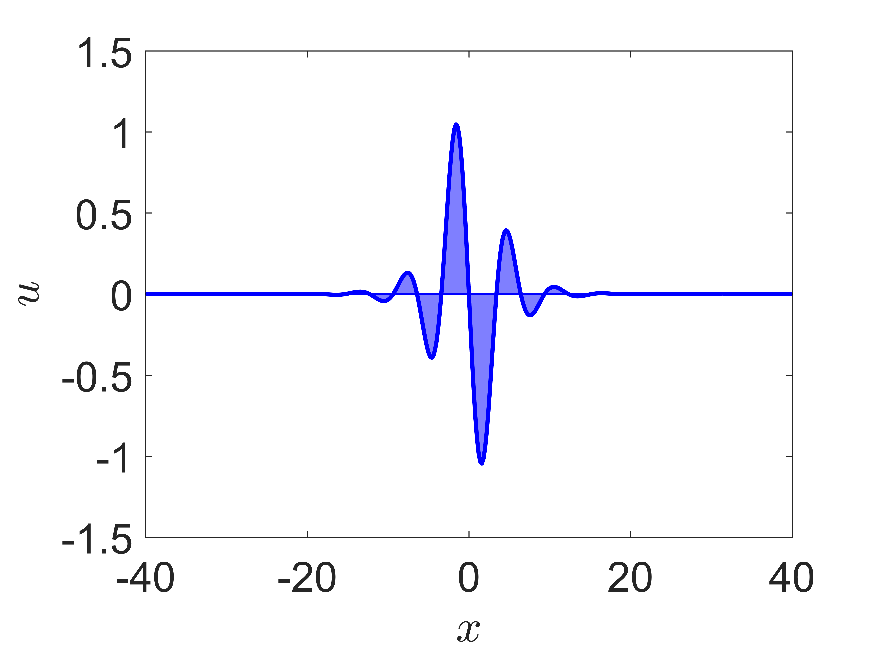}\label{subfig:prof_d}}\\
		\subfloat[]{\includegraphics[scale=0.19]{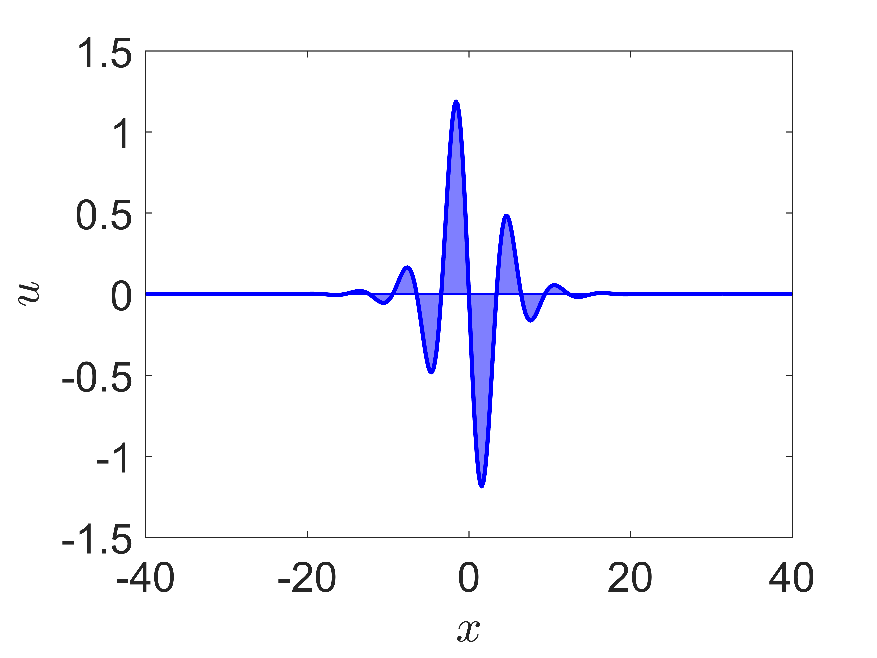}\label{subfig:prof_e}}\,\,
		\subfloat[]{\includegraphics[scale=0.19]{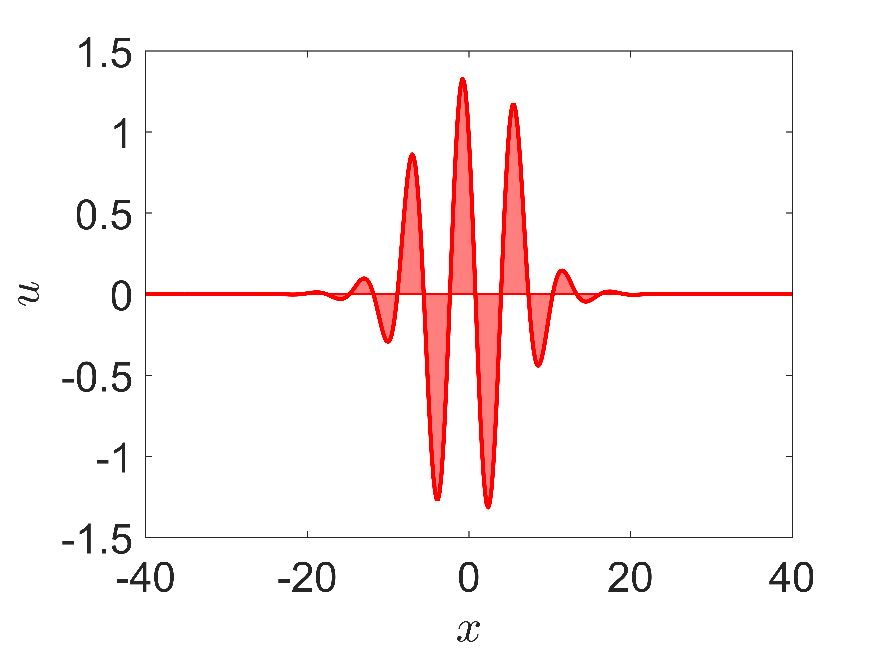}\label{subfig:prof_f}}
		\subfloat[]{\includegraphics[scale=0.19]{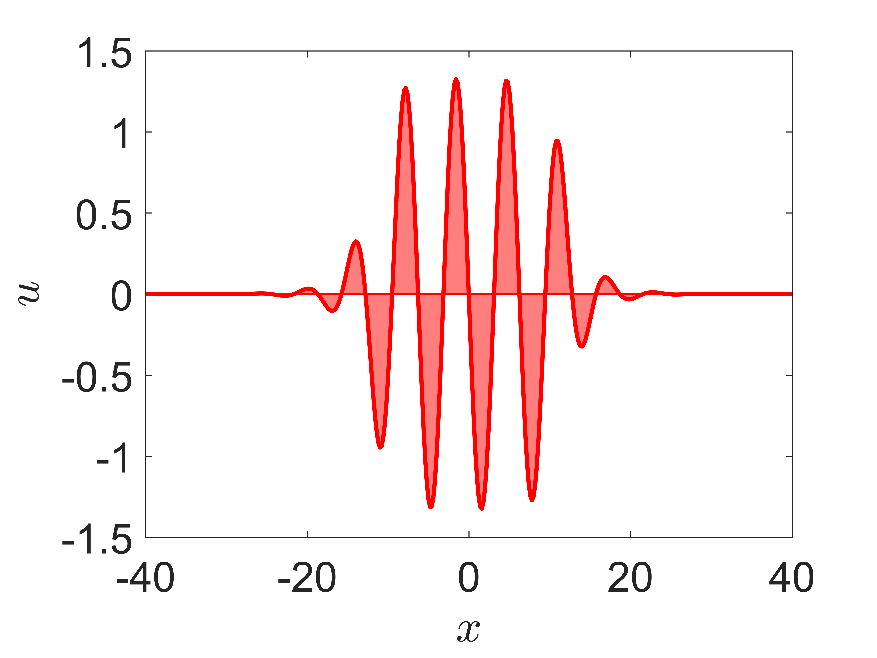}\label{subfig:prof_g}}\,\,
		\subfloat[]{\includegraphics[scale=0.19]{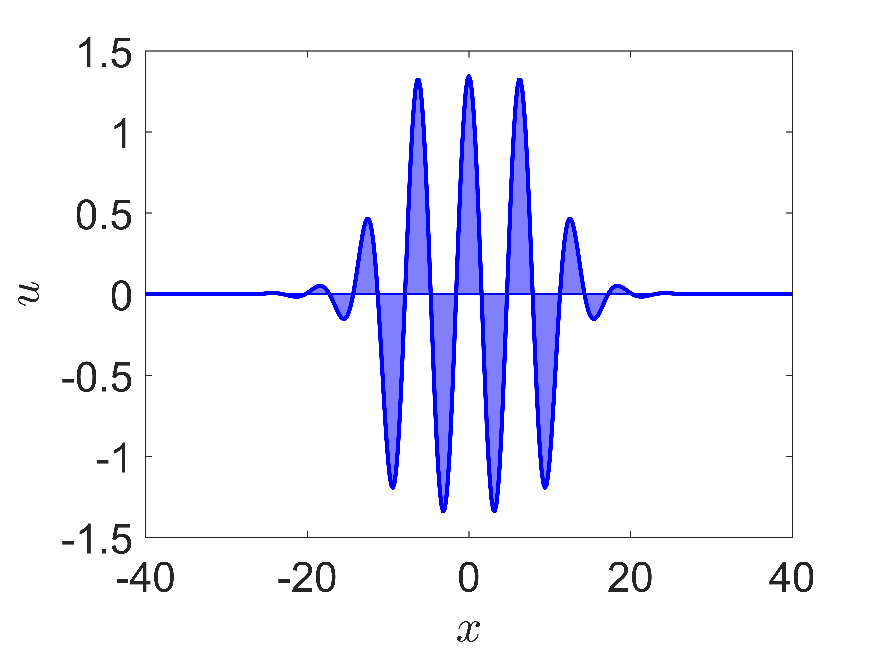}\label{subfig:prof_h}}
		\caption{Profile of localized solutions at the points indicated in Fig.\ \ref{fig:snakes}. {Symmetric: (a), (h); Anti-symmetric: (b), (d), (e), (g); Asymmetric: (c), (f).}}
		\label{fig:profiles_eigenvalues}
	\end{figure}
	
	\begin{figure}[tbhp!]
		\centering
		\subfloat[]{\includegraphics[scale=0.19]{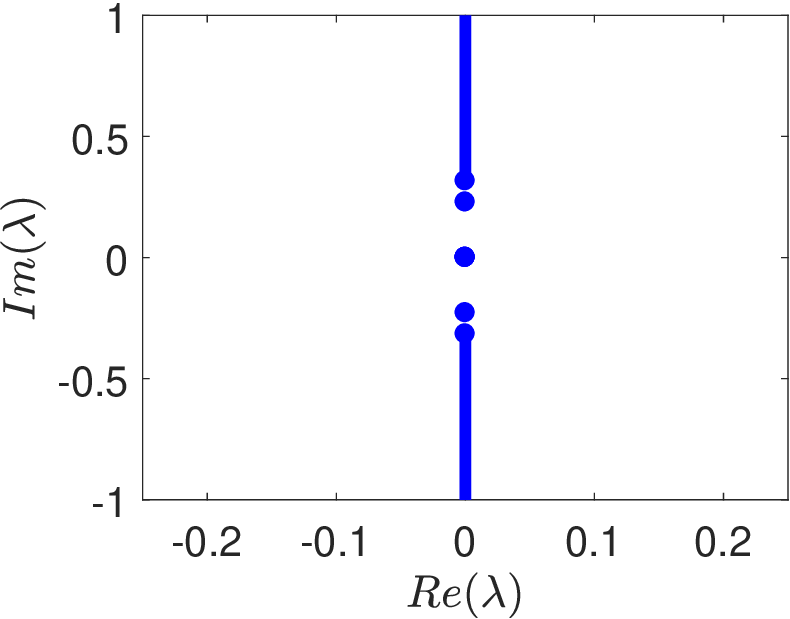}\label{subfig:eig_a}}\,\,
		\subfloat[]{\includegraphics[scale=0.19]{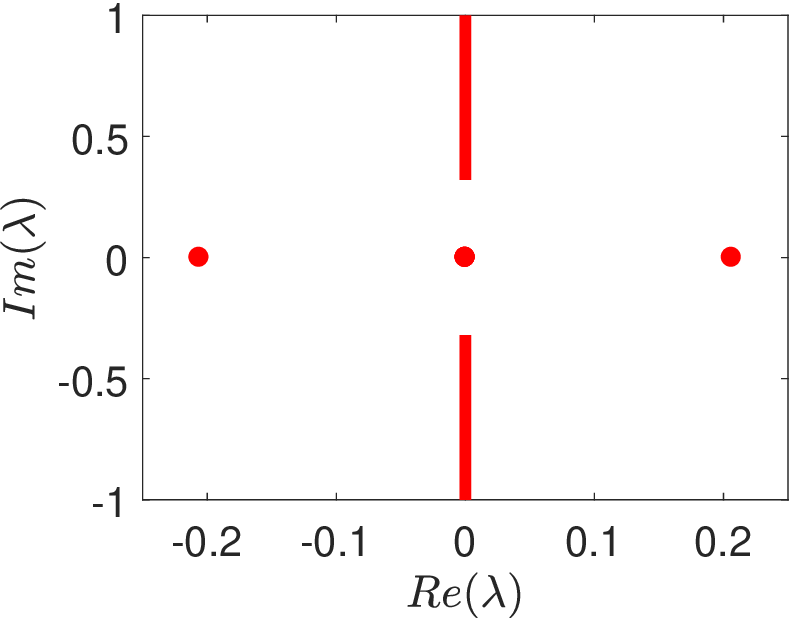}\label{subfig:eig_b}}\,\,
		\subfloat[]{\includegraphics[scale=0.19]{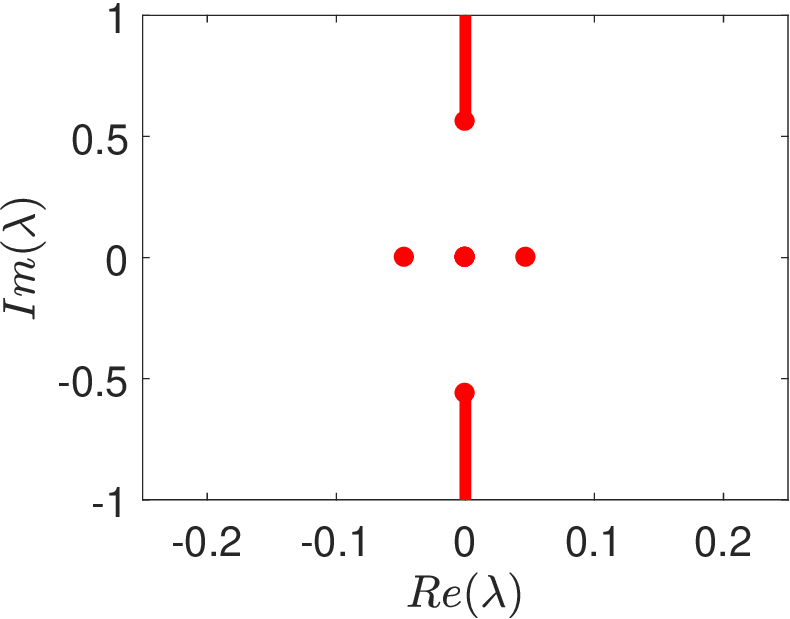}\label{subfig:eig_c}}\,\,
		\subfloat[]{\includegraphics[scale=0.19]{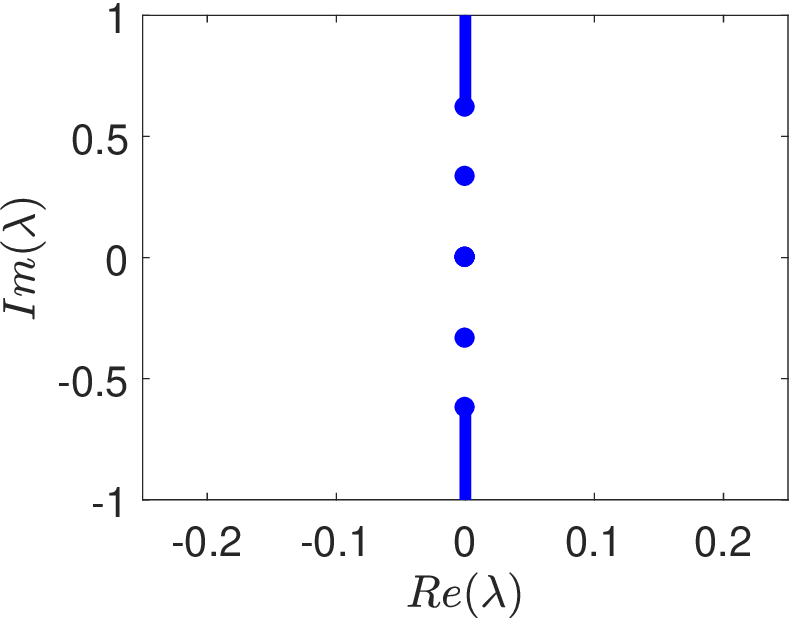}\label{subfig:eig_d}}\\
		\subfloat[]{\includegraphics[scale=0.19]{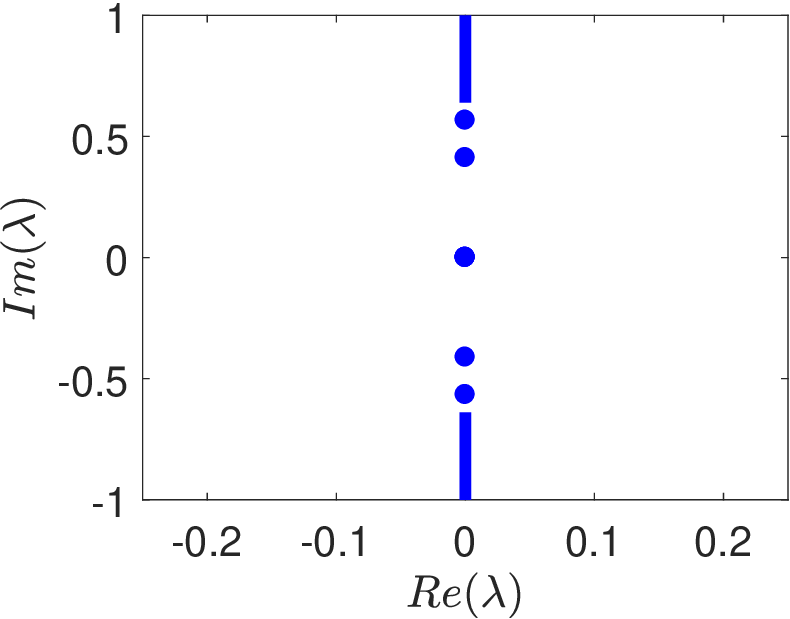}\label{subfig:eig_e}}\,\,
		\subfloat[]{\includegraphics[scale=0.19]{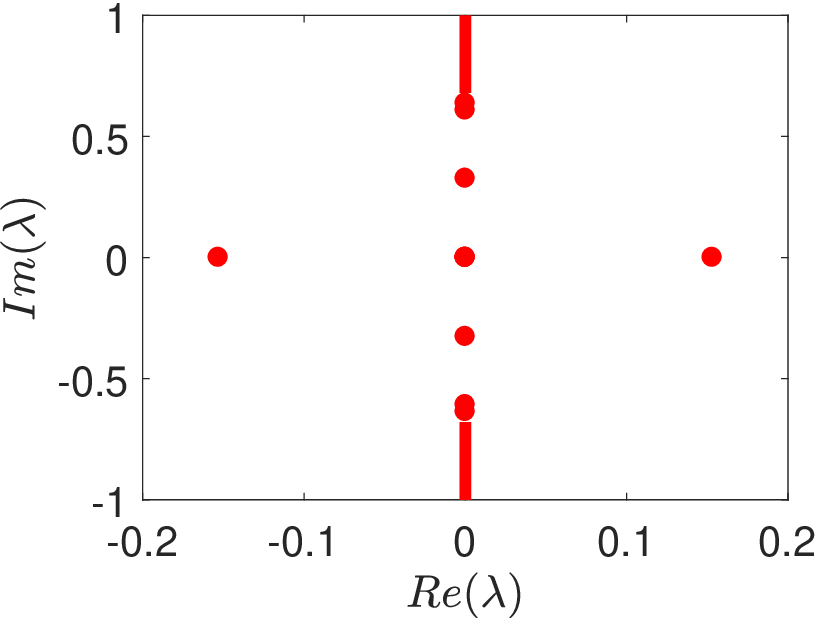}\label{subfig:eig_f}}\,\,
		\subfloat[]{\includegraphics[scale=0.19]{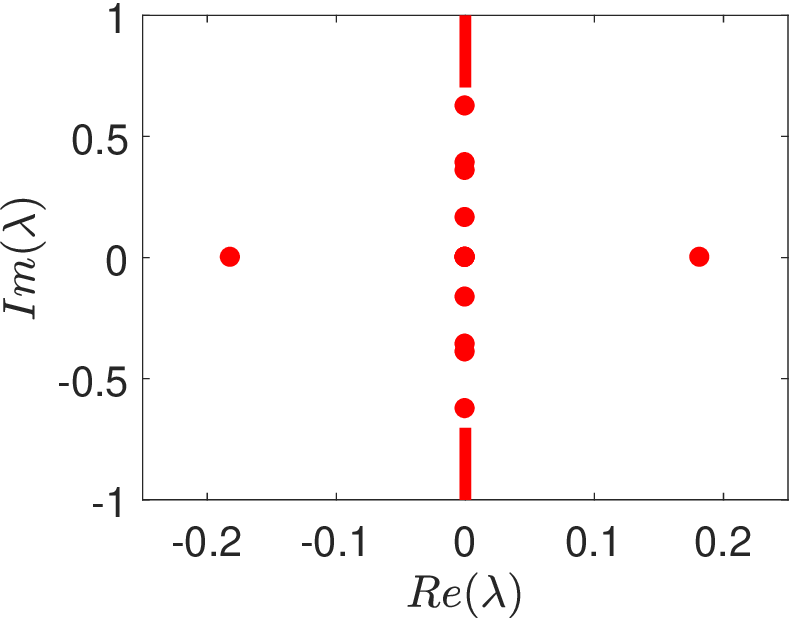}\label{subfig:eig_g}}\,\,
		\subfloat[]{\includegraphics[scale=0.19]{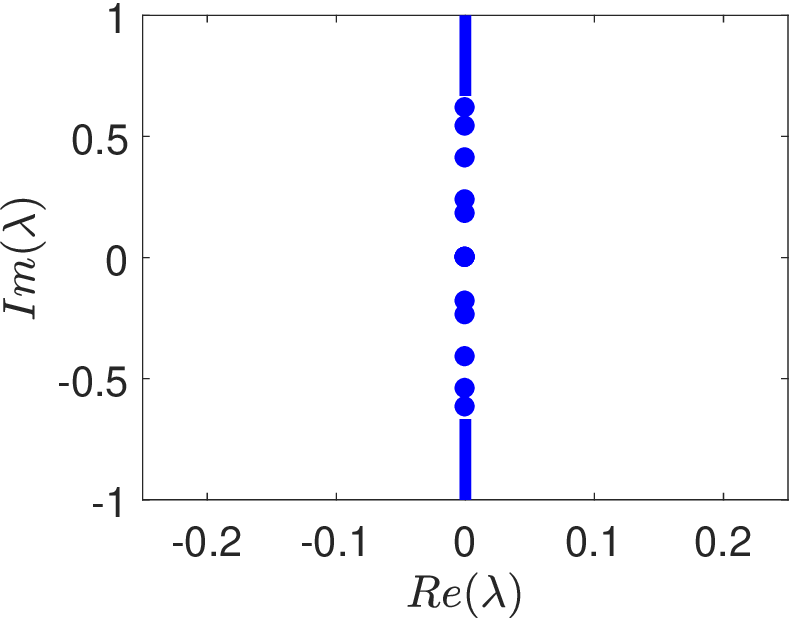}\label{subfig:eig_h}}
		\caption{{
				Spectrum of the localized solutions in Fig.\ \ref{fig:profiles_eigenvalues}.
		}}
		\label{fig:profiles_eigenvalues2}
	\end{figure}
	
	\section{Localised states} % and the Vakhitov-Kolokolov criterion}
\label{sec4}

\begin{figure}[htbp]
	\centering
	\subfloat[]{\includegraphics[scale=0.19]{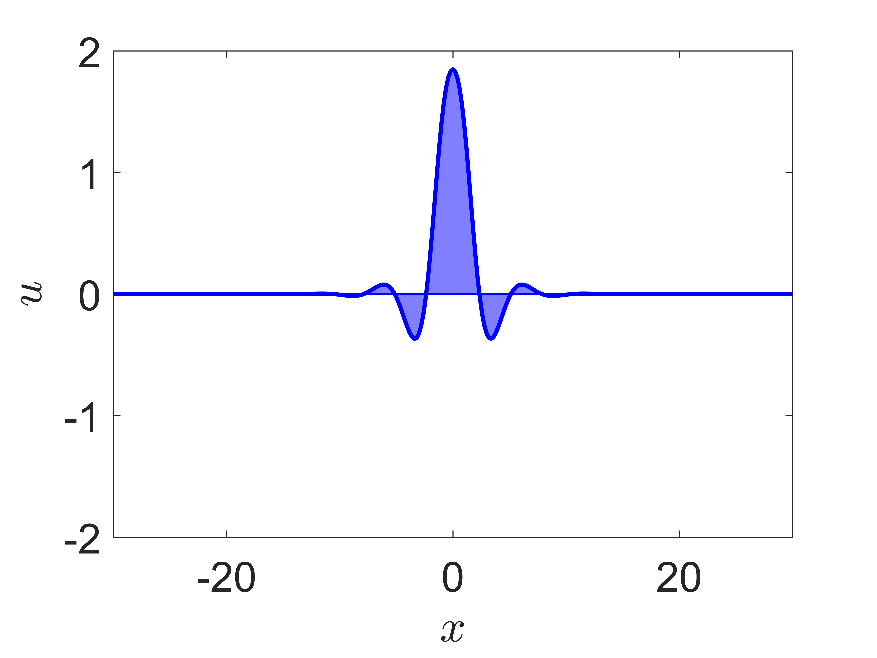}\label{subfig:prof_large_b3_a}}\,\,
	\subfloat[]{\includegraphics[scale=0.19]{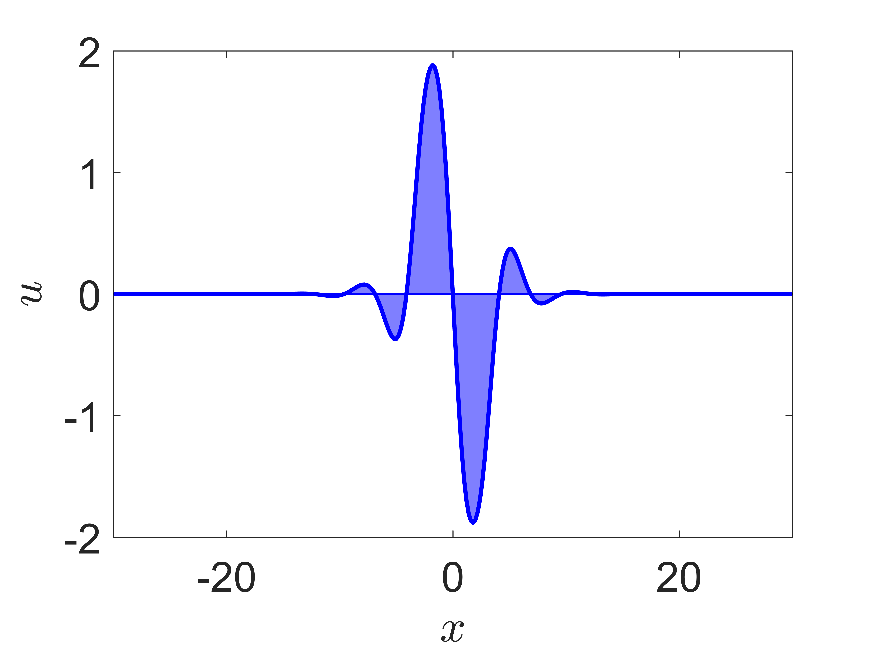}\label{subfig:prof_large_b3_b}}
	\subfloat[]{\includegraphics[scale=0.19]{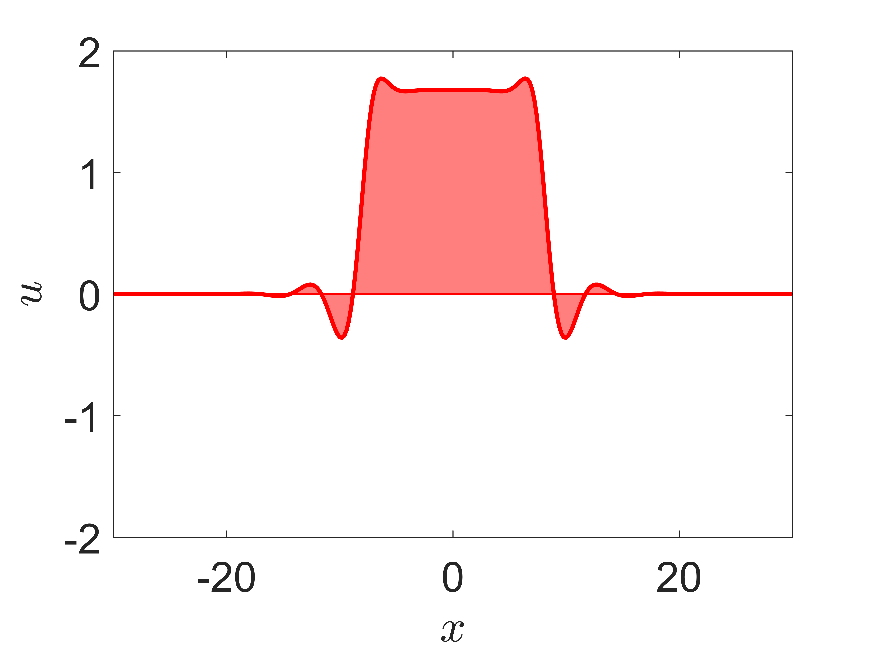}\label{subfig:prof_large_b3_c}}\,\,
	\subfloat[]{\includegraphics[scale=0.19]{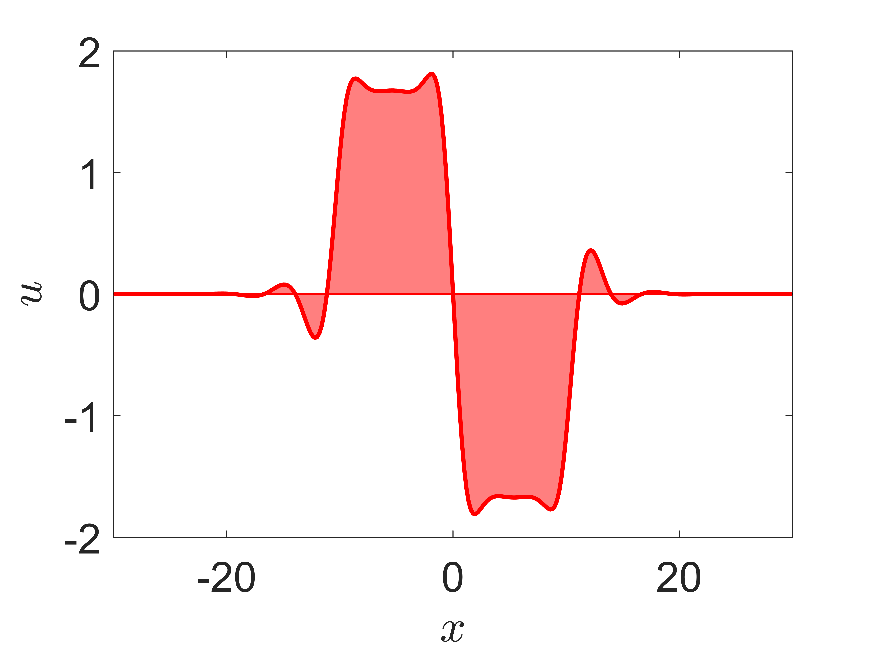}\label{subfig:prof_large_b3_d}}
	\caption{Localised solution profiles of the CSHE for large $b_3$, i.e., $3.75$, that correspond to the points indicated in Fig.\ \ref{fig:snakes_b3}.
	}
	\label{fig:prof_eigen_b3}
\end{figure}

\begin{figure}[htbp]
	\centering
	\subfloat[]{\includegraphics[scale=0.19]{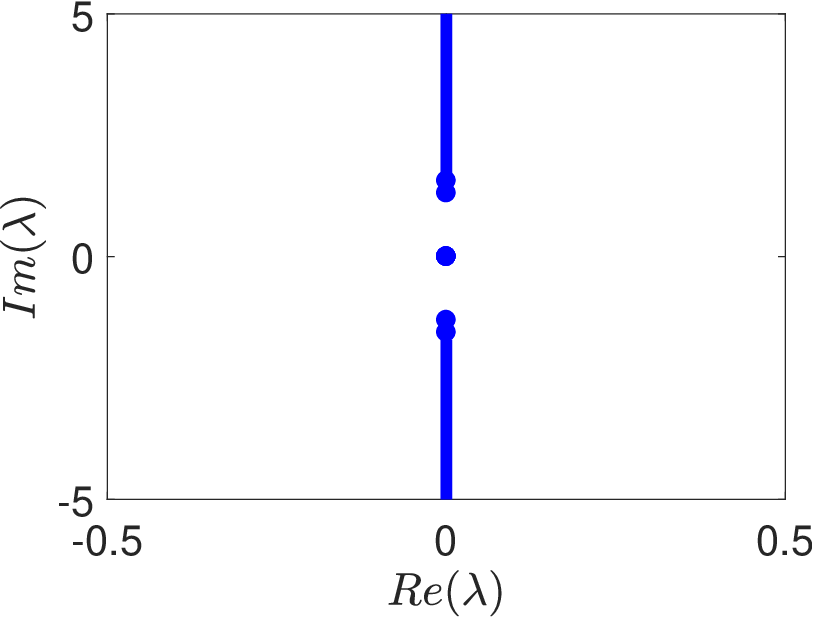}\label{subfig:eig_large_b3_a}}\,\,
	\subfloat[]{\includegraphics[scale=0.19]{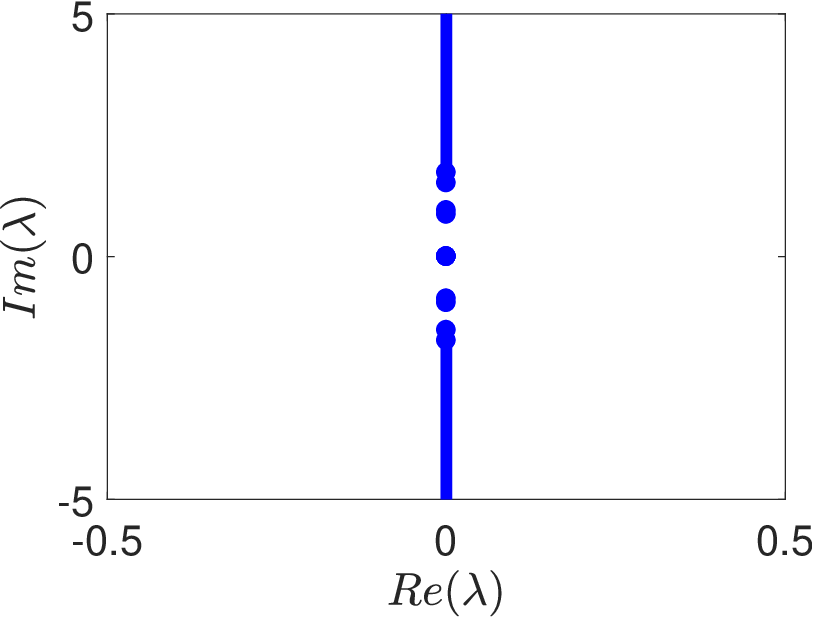}\label{subfig:eig_large_b3_b}}\,\,
	\subfloat[]{\includegraphics[scale=0.19]{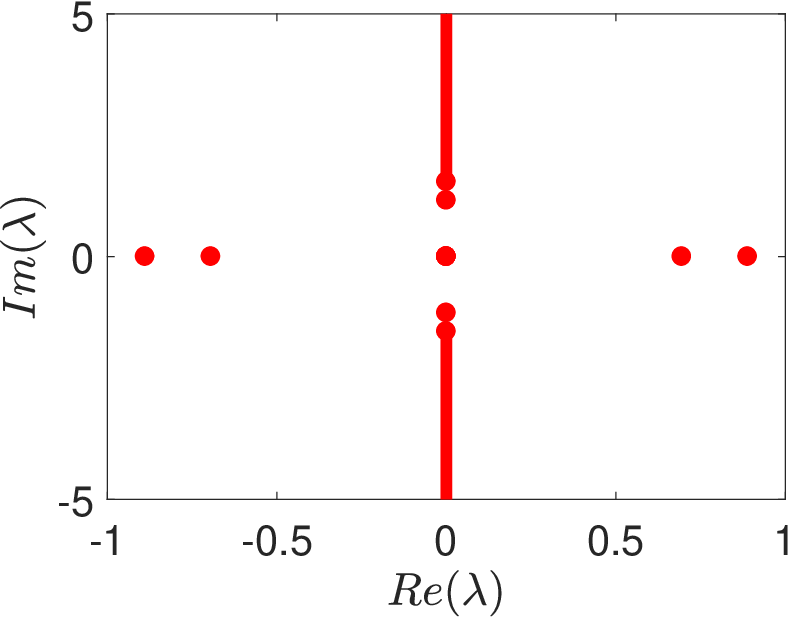}\label{subfig:eig_large_b3_c}}\,\,
	\subfloat[]{\includegraphics[scale=0.19]{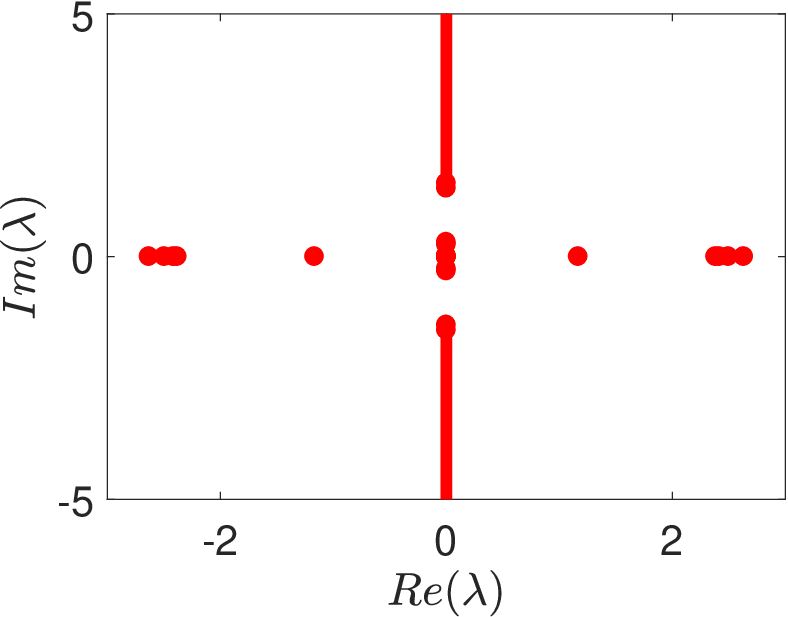}\label{subfig:eig_large_b3_d}}
	\caption{Spectrum of the solutions in Fig.\ \ref{fig:prof_eigen_b3}.
	}
	\label{fig:prof_eigen_b32}
\end{figure}

The steady-state equation \eqref{gov2} supports localized states that bifurcate from \(\omega = 0\) \cite{burke2007snakes}. Near the bifurcation point, the method of multiple scales yields an approximate solution \cite{burke2007snakes}:
\begin{equation}
	u_l(x) = 2\left(\frac{2\omega}{3b_3}\right)^{\frac{1}{2}} \sech\left(\frac{x\sqrt{\omega}}{2}\right)\cos(x + \phi),
	\label{sol}
\end{equation}
where the phase shift \(\phi = 0,\, \pi\) corresponds to symmetric solutions, and \(\phi = \pi/2,\, 3\pi/2\) corresponds to anti-symmetric solutions. Numerical continuation of \eqref{sol}, with stability indicated by color coding, is shown in Fig.\ \ref{fig:add1}, where the vertical axis represents the solution norm defined in Eq.\ \eqref{P}.

The bifurcation diagram of localized solutions in Fig.\ \ref{fig:snakes} reveals the characteristic oscillatory structure known as homoclinic snaking \cite{woods1999heteroclinic}. This structure arises from homoclinic orbits in phase space and features multiple hysteresis. The two snaking branches are connected by rungs representing asymmetric solutions. The mechanisms driving the snaking phenomenon, including the determination of the snaking region's width, have been extensively studied, as seen in \cite{avitabile2010snake,coullet2000stable,dean2011exponential,kozyreff2006asymptotics,susanto2011variational,matthews2011variational}.

While the overall structure of the bifurcation diagram resembles that of the SHE \eqref{dshe} \cite[Fig.\ 4]{burke2007snakes}, significant differences emerge in the stability of solutions. In the real SHE, symmetric and {anti-symmetric} localized states are unstable along the lower parts of the bifurcation diagram. In contrast, for the conservative CSHE studied here, symmetric states bifurcate as stable solutions (see the inset of Fig.\ \ref{fig:snakes}) but lose stability before reaching the first turning point. Conversely, asymmetric states bifurcate as unstable solutions but stabilize before the first turning point.

Several solution profiles corresponding to the marked points in Fig.\ \ref{fig:snakes} are shown in Fig.\ \ref{fig:profiles_eigenvalues}. As the continuation progresses away from the bifurcation point \(\omega = 0\), the localized solutions broaden, forming profiles that consist of two fronts connecting the zero homogeneous solution to the upper periodic state (see Fig.\ \ref{fig:unisol_persol}). The eigenvalue spectra of these solutions, depicted in Fig.\ \ref{fig:profiles_eigenvalues2}, reveal that instabilities, when present, are driven by a finite number of real eigenvalues. Notably, the region of homoclinic snaking lies within the stability region of the upper periodic state (Fig.\ \ref{fig:unisol_persol}), ensuring that localized solutions with extended periodic parts can remain stable. Instabilities, when they occur, typically arise from the manner in which the fronts connect to the periodic solution.

Burke and Knobloch \cite{burke2007snakes} observed that the snaking profile in the bifurcation diagram of localized solutions vanishes for sufficiently large values of \(b_3\). Figure \ref{fig:snakes_b3} illustrates this phenomenon for \(b_3 = 3.75\), where snaking is absent. Instead, localized states consist of two fronts connecting uniform solutions. In this regime, solutions higher up in the bifurcation diagram are entirely unstable. This contrasts sharply with the dissipative SHE, where localized solutions alternate between stable and unstable branches, as shown in \cite[Fig.\ 10]{burke2007snakes}. This difference can be attributed to the instability of the upper uniform solution \(u_+\) (see Fig.\ \ref{fig:unisol_persol}), which forms the plateau of these localized states. Examples of localized solutions and their corresponding eigenvalue spectra for this regime are presented in Fig.\ \ref{fig:prof_eigen_b3} and Fig.\ \ref{fig:prof_eigen_b32}.

\begin{figure}[h!]
	\centering
	\subfloat[]{\includegraphics[scale=0.39]{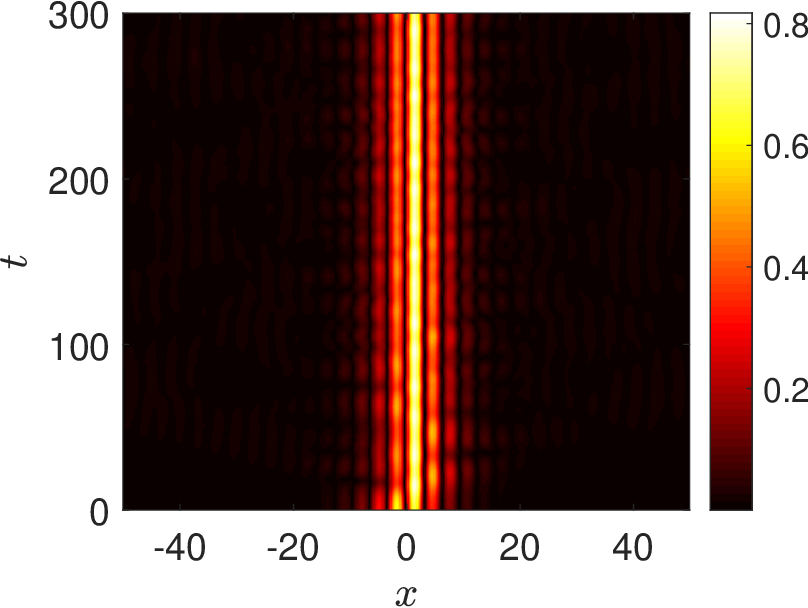}\label{subfig:time_dynamics_b}}\,
	\subfloat[]{\includegraphics[scale=0.39]{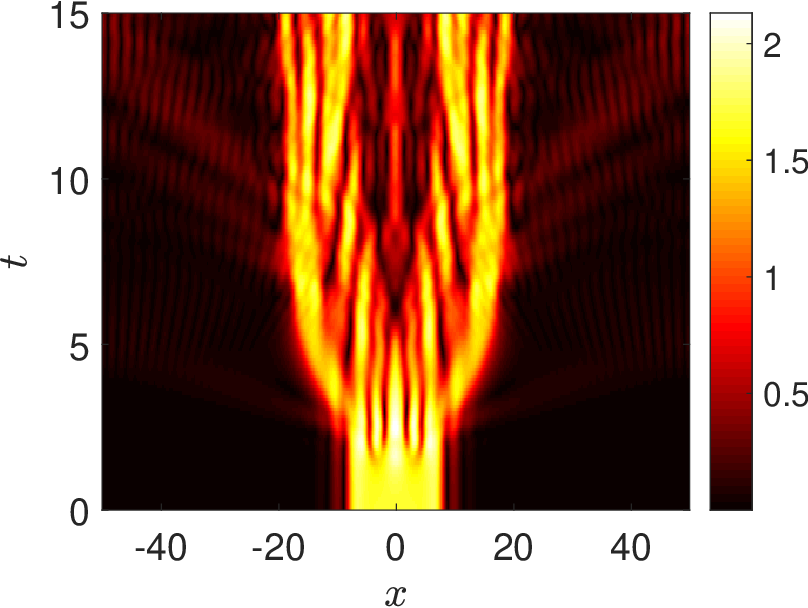}\label{subfig:time_dynamics_c_b3}}
	%	\subfloat[]{\includegraphics[scale=0.5]{time_dynamics_c_b3_k=0_5}\label{subfig:time_dynamics_c_b3_k=0.5}}
	\caption{%Time-dynamics point (c)
		Typical time dynamics of a localized solution when it is unstable. (a) Time evolution of the unstable solution at point (b) in Fig.\ \ref{fig:snakes}. (b) Time evolution of the unstable solution at point (c) in Fig.\ \ref{fig:snakes_b3}. 
	}
	\label{fig:time_dynamics_c_b3}
\end{figure}

\begin{figure}[tbhp!]
	\centering
	\subfloat[$L_+$ and $\phi=0,\pi$]{\includegraphics[scale=0.39]{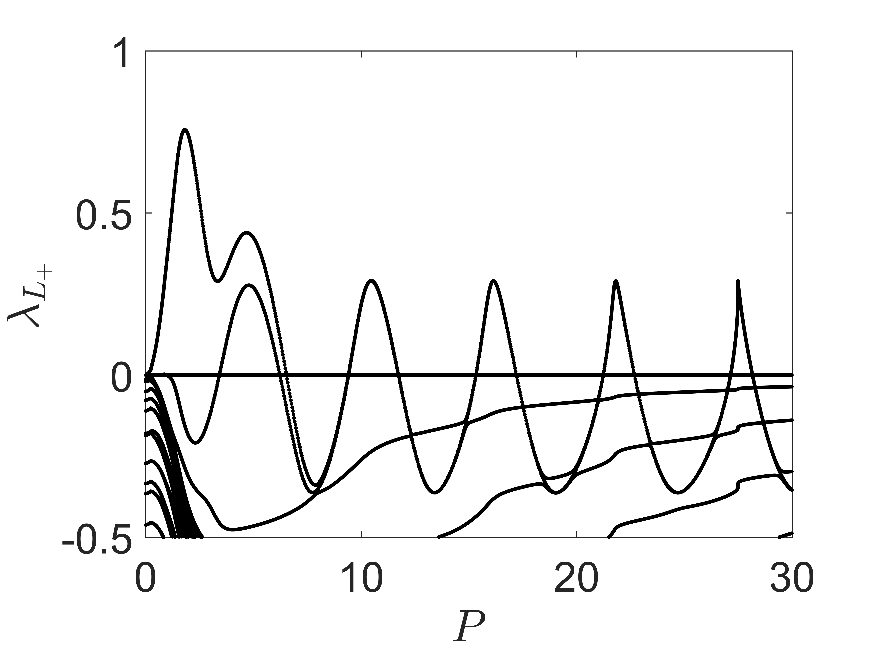}\label{subfig:Lplus_0}}\,\,
	\subfloat[$L_+$ and $\phi=\pi/2,3\pi/2$]{\includegraphics[scale=0.39]{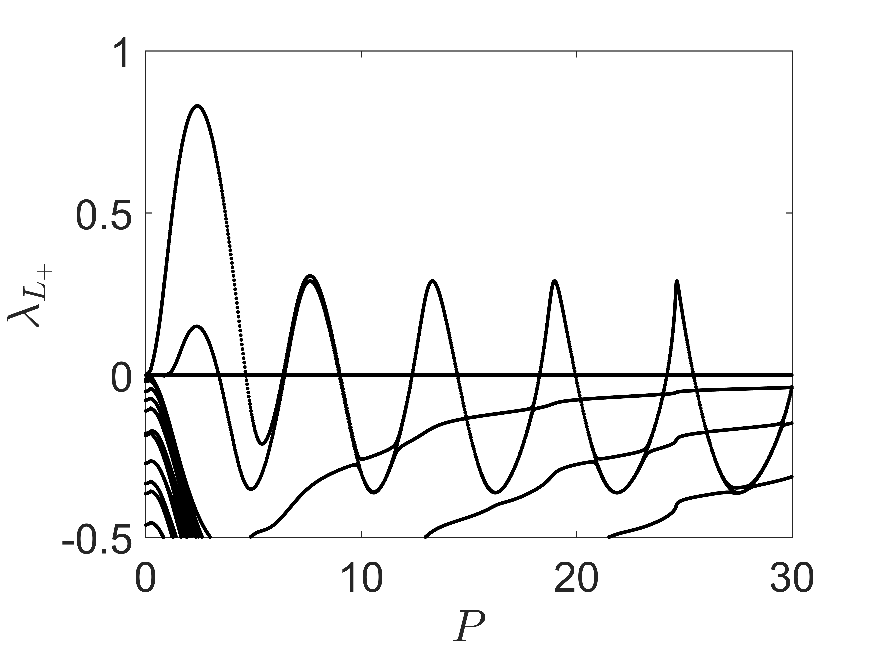}\label{subfig:Lplus_1}}\\
	\subfloat[$L_-$ and $\phi=0,\pi$]{\includegraphics[scale=0.39]{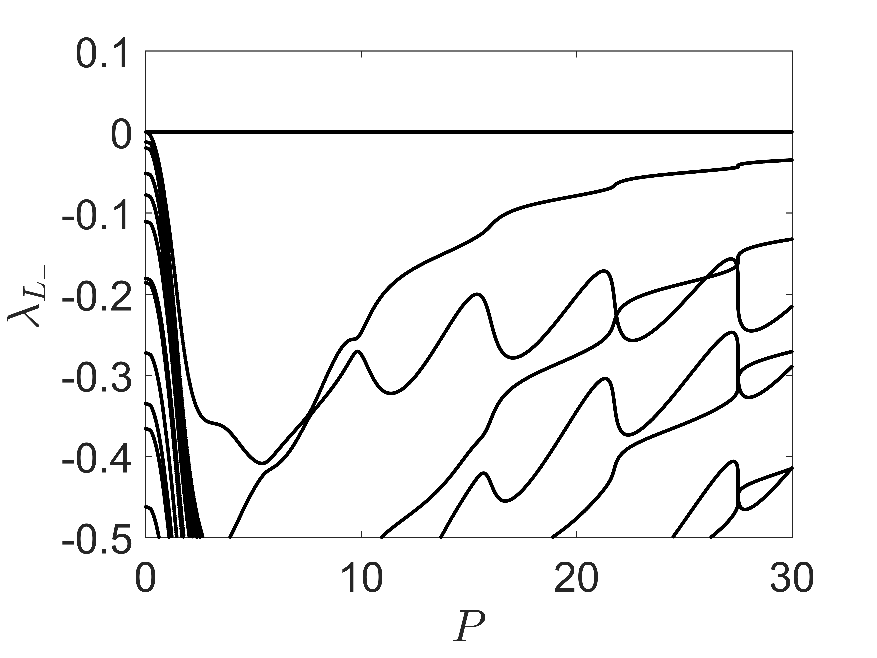}\label{subfig:Lmin_0}}\,\,
	\subfloat[$L_-$ and $\phi=\pi/2,3\pi/2$]{\includegraphics[scale=0.39]{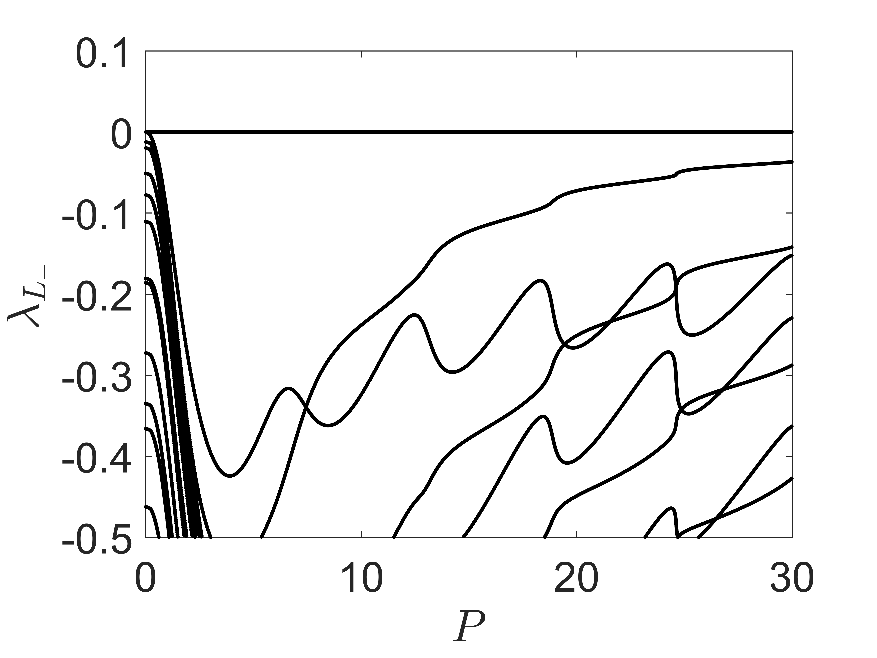}\label{subfig:Lmin_1}}
	\caption{{Spectrum $\lambda$ of the linear operator $L_+$ and $L_-$ for 
			phase shifts $\phi=0,\pi$ and $\pi/2,3\pi/2$ as a function of the power $P$ for the localized states along the bifurcation diagram in Fig.\ \ref{fig:snakes}.}
	}
	\label{fig:LL}
\end{figure}

Finally, we examine the typical time dynamics of unstable localized solutions in the system. Figure \ref{fig:time_dynamics_c_b3} illustrates the time evolution of two such solutions: the left panel corresponds to the unstable anti-symmetric solution at point (b) in Fig.\ \ref{fig:snakes}, while the right panel shows the dynamics of the flat-top state at point (c) in Fig.\ \ref{fig:snakes_b3}.

In the first case, the unstable anti-symmetric solution transitions into a stable symmetric state. This behavior is representative of a general dynamic observed across unstable localized solutions within the homoclinic snaking diagram: instability drives a reconfiguration of the solution profile, ultimately stabilizing into a symmetric configuration. This symmetry-restoring process underscores the system's inherent preference for energetically favorable symmetric solutions, which remain stable within the snaking structure.

In the second case, the right panel reveals the time dynamics of an unstable flat-top state. Here, the localized wave structure gradually disintegrates, leading to the dissolution of the plateau. This behavior aligns with the instability of the upper uniform state \(u_+\) observed in Fig.\ \ref{subfig:unisol_upper_d}, where perturbations grow and disrupt the solution's uniformity. The similarity between the dynamics of the flat-top localized state and the unstable upper uniform state suggests that the plateau acts as a transitional structure linking localized and uniform states.

These examples highlight the diversity of instabilities in the system: while some unstable localized solutions evolve into stable configurations, others are inherently unstable and decay completely. The interplay between localized wave structures, their stability, and their dynamic transformations shows the rich behavior of the system \eqref{gov}, particularly within the context of homoclinic snaking and beyond.

\section{Vakhitov-Kolokolov stability criterion}
\label{sec5}

In this section, we analyze the (in)stability of localized solutions reported in Section \ref{sec4}. We establish a connection between the stability results for the real dissipative SHE, obtained by numerically computing the spectrum of the operator \( L_+ \) \cite{burke2007snakes}, and the current problem in the context of the Schr\"odinger equation. Mathematically, we relate the eigenvalues of \( \mathcal{J}\mathcal{L} \) in Eq.\ \eqref{stab} to those of \( L_\pm \), particularly focusing on the slope of the bifurcation diagrams in Fig.\ \ref{fig:add1}.

A remarkable connection exists between the eigenvalues of the operator \( \mathcal{J}\mathcal{L} \) and those of \( \mathcal{L} \), which has been extensively studied for the general nonlinear Schr\"odinger equation. This analysis originated with the discovery of the relationship between the linear stability of a localized solution and the slope of its power curve by \cite{vakhitov1973stationary}, building on earlier work by \cite{zakharov1968instability}. This result is now known as the Vakhitov-Kolokolov stability criterion. Grillakis et al.\ \cite{grillakis1987stability,grillakis1990stability} later generalized this criterion to abstract Hamiltonian systems, demonstrating that it also guarantees the orbital stability of localized states.

Let \( u(x) \) be a localized solution in the bifurcation diagram in Fig.\ \ref{fig:add1}, and let \( n_+(L_\pm) \) denote the number of positive discrete eigenvalues of the operator \( L_\pm \). Note that \( L_-u(x) = 0 \) is simply Eq.\ \eqref{gov2}, and taking its derivative with respect to \( x \) yields \( L_+u_x = 0 \). This implies that \( 0 \) is always an eigenvalue of \( L_\pm \). If there were no fourth-order derivative in \( L_\pm \), we could use the Sturm-Liouville theorem \cite{amrein2005sturm} to determine \( n_+(L_\pm) \). However, in the presence of higher-order derivatives, while we can still order the eigenvalues, there is no longer a direct relationship between an eigenvalue index and the number of zeros of the corresponding eigenfunction \cite{amrein2005sturm}. Consequently, in our case, we rely on numerical results to assert that zero is the largest eigenvalue of \( L_- \), i.e., \( n_+(L_-) = 0 \). With this, we state the following theorem.

\begin{thm}\label{theorem}
	For a localized state of the conservative CSHE \eqref{gov} with \( n_+(L_-) = 0 \),
	\begin{itemize}
		\item[(i)] if \( n_+(L_+) = 0 \), the localized state is linearly stable;
		\item[(ii)] if \( n_+(L_+) = 1 \), the localized state is linearly unstable if and only if \( P'(\omega) < 0 \);
		\item[(iii)] if \( n_+(L_+) \geq 2 \), the localized state is linearly unstable.
	\end{itemize}
	Moreover, in the case of instability, there will be \( n_+(L_+) \) real unstable eigenvalues if \( P'(\omega) < 0 \) and \( n_+(L_+) - 1 \) real unstable eigenvalues if \( P'(\omega) > 0 \).
\end{thm}

Parts (i--iii) of Theorem \ref{theorem} are adapted from \cite[Theorem 5.2]{yang2010nonlinear} for multidimensional NLS equations (i.e., with second derivatives in space) with general nonlinearities and external potentials. The argument in our case follows straightforwardly. The final statement in the theorem is proven in \cite{grillakis1988linearized} and further discussed in, e.g., \cite[Theorem 1 and Remark 6.1]{chugunova2010count} and \cite[Remark 3.1]{kapitula2004counting}.

In Fig.\ \ref{fig:LL}, we show the spectrum of the linear operators \( L_+ \) and \( L_- \) for phase shifts \( \phi = 0, \pi \) and \( \phi = \pi/2, 3\pi/2 \) as a function of the power \( P \) for the localized states along the bifurcation diagram in Fig.\ \ref{fig:snakes}. The spectrum of \( L_- \) is always negative, i.e., \( n_+(L_-) = 0 \), while the number of positive eigenvalues of \( L_+ \) varies between \( 0 \) and \( 2 \).

Near the bifurcation point \( \omega = 0 \), we find that \( n_+(L_+) = 1 \) for the symmetric localized state and \( n_+(L_+) = 2 \) for the anti-symmetric one. Consequently, the bifurcating localized states are unstable in the real SHE but not necessarily in the complex equation. Using part (ii) of Theorem \ref{theorem}, we conclude that the symmetric solution is stable, which agrees with our numerical results. Before the first turning point along the lowest bifurcation curves, we have \( n_+(L_+) = 2 \) for symmetric states and \( n_+(L_+) = 1 \) for anti-symmetric ones, explaining the stability changes in Fig.\ \ref{fig:snakes}. Further up in the snaking curves, corresponding to large solution norms, we generally find either \( n_+(L_+) \geq 2 \) or \( n_+(L_+) = 0 \). As a result, localized solutions with well-developed periodic states typically share the same stability properties in both the real and complex SHEs.

Finally, for localized solutions along the bifurcation diagram in Fig.\ \ref{fig:snakes_b3}, we can also use Theorem \ref{theorem} to establish their instability, particularly in the collapsed snaking region. Since the plateau of the localized state (see Fig.\ \ref{fig:prof_eigen_b3}) tends toward the unstable upper uniform state \( u_+ \) (see Fig.\ \ref{fig:unisol_persol}), it is natural to expect \( n_+(L_+) \geq 2 \) in that region, implying instability of the localized states.

\section{Conclusion}
\label{conc}	

In this study, we have investigated the conservative CSHE with cubic and quintic nonlinearities, focusing on its homogeneous, spatially periodic, and localized solutions. While the existence of such solutions has been well-established for the real SHE, our novel contribution lies in analyzing their stability within the context of the complex model. This distinction is significant, as the conservative complex equation introduces symplectic structures and additional constraints that fundamentally alter stability characteristics compared to its real-valued counterpart.

A key outcome of our work is the development of a generalized Vakhitov-Kolokolov criterion, which bridges the stability properties of localized solutions in the SHE and CSHE. This theoretical framework provides valuable insights into the stability transitions and bifurcation structures of the system, offering a deeper understanding of the interplay between conserved quantities and solution dynamics. Our findings mark a foundational step toward the systematic classification of various solutions in the CSHE and set the stage for exploring more generalized and higher-dimensional models.

Looking ahead, several intriguing directions for future research emerge. A natural extension of this work would involve exploring the fully complex SHE by introducing parameters such as complex-valued \(\omega\) and \(b_3\). This generalization could reveal richer dynamics, including complex bifurcation structures and new classes of localized states. Additionally, investigating the long-term dynamics of homogeneous states and spatially periodic solutions under modulational instability could shed light on the formation of rogue waves—solutions that are highly localized in both space and time \cite{toenger2015emergent}. The emergence of such phenomena in the CSHE, driven by modulational instability, represents an interesting topic for future exploration.

Furthermore, extending the analysis to higher-dimensional systems offers the potential to uncover novel patterns and stability behaviors unique to multidimensional settings. Another promising direction involves incorporating dissipative terms into the conservative framework to study the transition between conservative and dissipative dynamics. %This approach could bridge the gap between theoretical predictions and experimental observations in physical systems such as nonlinear optics, fluid dynamics, and reaction-diffusion systems.

\begin{acknowledgements}
	RK acknowledges financial support from Riset PPMI KK FMIPA ITB (3D/IT1.C02/KU/2025). 	HS acknowledged support by Khalifa University through a Competitive Internal Research Awards Grant (No.\ 8474000413/CIRA-2021-065) and Research \& Innovation Grants (No.\ 8474000617/RIG-S-2023-031 and No.\ 8474000789/RIG-S-2024-070).
\end{acknowledgements}

%\section*{Conflict of interest}
%The authors declare that they have no conflict of interest.
%

\section*{Compliance with ethical standards}
\subsection*{Conflict of interest} The authors declare no conflict of interest.

%\bibliographystyle{elsarticle-num}
%%\bibliographystyle{abbrv}
%\bibliography{shnls2}

\end{document}